\documentclass{article}

\usepackage{PRIMEarxiv}

\usepackage[utf8]{inputenc} 
\usepackage[T1]{fontenc}    
\usepackage{hyperref}       
\usepackage{url}            
\usepackage{booktabs}       
\usepackage{amsfonts}       
\usepackage{nicefrac}       
\usepackage{microtype}      
\usepackage{lipsum}
\usepackage{fancyhdr}       
\usepackage{graphicx}       
\usepackage{array}
\graphicspath{{media/}}     

\title{Evaluation-Driven Development and Operations of LLM Agents: A Process Model and Reference Architecture}

\author{
Boming Xia\textsuperscript{1,2,3},
Qinghua Lu\textsuperscript{3,4,1},
Liming Zhu\textsuperscript{3,4},
Zhenchang Xing\textsuperscript{3},
Dehai Zhao\textsuperscript{3},
Hao Zhang\textsuperscript{3,1}
\\[3pt]
\textsuperscript{1} Responsible AI Research Centre, Australia\\
\textsuperscript{2} Adelaide University, Australia\\
\textsuperscript{3} CSIRO's Data61, Australia\\
\textsuperscript{4} University of New South Wales, Australia
}

\begin{document}
\maketitle

\begin{abstract}
Large Language Models (LLMs) have enabled the emergence of LLM agents, systems capable of pursuing under-specified goals and adapting after deployment. Evaluating such agents is challenging because their behavior is open ended, probabilistic, and shaped by system-level interactions over time. Traditional evaluation methods, built around fixed benchmarks and static test suites, fail to capture emergent behaviors or support continuous adaptation across the lifecycle. To ground a more systematic approach, we conduct a multivocal literature review (MLR) synthesizing academic and industrial evaluation practices. The findings directly inform two empirically derived artifacts: a process model and a reference architecture that embed evaluation as a continuous, governing function rather than a terminal checkpoint. Together they constitute the evaluation-driven development and operations (EDDOps) approach, which unifies offline (development-time) and online (runtime) evaluation within a closed feedback loop. By making evaluation evidence drive both runtime adaptation and governed redevelopment, EDDOps supports safer, more traceable evolution of LLM agents aligned with changing objectives, user needs, and governance constraints.
\end{abstract}


\section{Introduction}
\label{sec:intro}

Recent advances in large language models (LLMs) have led to the emergence of LLM-based agents (LLM agents), which are capable of autonomously performing complex tasks (e.g., AI Scientist~\cite{lu2024ai}). Unlike traditional AI/LLM systems that rely on predefined inputs and structured task execution, LLM agents dynamically interpret high-level, under-specified user goals. They perceive context, reason, plan, and execute workflows, often integrating external tools, knowledge bases, and other agents to extend their capabilities~\cite{bass25engineering}. In this sense, LLMs act as foundational components within broader socio-technical systems in which capabilities and behaviors arise from interactions among models, tools, data, and human or organizational environments.

Despite growing adoption, LLM agents raise concerns about performance and safety~\cite{lu2023responsible}. Their open-ended reasoning, planning, and execution can lead to degraded task performance (e.g., goal drift, tool misuse, or reasoning errors) and safety issues (e.g., harmful outputs, privacy breaches, or regulatory violations), particularly under ambiguity, in novel scenarios, or within evolving contexts. Moreover, autonomy in selecting tools, adjusting workflows, and adapting responses raises governance challenges when actions diverge from human intent or regulatory constraints. Rigorous, continuous evaluation is therefore essential to ensure that LLM agents operate as expected and evolve safely over time.
In this paper, \textit{evaluation means applying certain criteria to observed outcomes to judge acceptability relative to stated objectives; it serves as an umbrella term that subsumes testing, benchmarking, validation/verification, online monitoring, and risk/impact assessment}~\cite{xia2024ai}.

While research on benchmarking and testing frameworks for LLMs and LLM agents is expanding (e.g.,~\cite{Li2024Survey,liuagentbench,kapoor2025holistic}), most existing approaches focus on model-level behavior, standardized benchmarks, static datasets, and aggregate scores. Such frameworks provide limited visibility into intermediate artifacts and long-horizon workflows, and they rarely support fine-grained analyses of specific task slices, user cohorts, or operational contexts to detect performance drift or emerging risks. They are also seldom connected to runtime observability systems that collect behavioral traces, or to policy mechanisms that enforce safety or compliance constraints. As a result, they fall short of addressing the full scope of system-level evaluation during operation~\cite{shreya2025indefense}.

Classical software engineering methods such as Test-Driven Development (TDD) and Behavior-Driven Development (BDD) also face limitations in this setting. These methods assume stable specifications, executable oracles, and predominantly pre-deployment test phases. In contrast, LLM agents are non-deterministic, pursue under-specified goals, and continue to adapt after deployment. Their outcomes may be graded or comparative, and adjudication can require pluggable judges or human oversight. Consequently, neither existing benchmarks and testing frameworks nor TDD/BDD alone can provide the continuous, lifecycle-wide evaluation and governance that LLM agents require. To achieve this, effective evaluation must consider the following:

\begin{itemize}
    \item \textbf{Evaluating LLM agents at the system level.}
    Current evaluation frameworks predominantly assess LLMs at the model level, focusing on prompt–response correctness or benchmark-based scores. However, LLM agents are compound AI systems that include not only an LLM but also non-model components~\cite{lu2024towards}, such as context engines, planning modules, memory, and guardrails~\cite{shamsujjoha2024taxonomymultilayeredruntimeguardrails}. While tools and frameworks like Inspect.AI\footnote{\url{https://github.com/UKGovernmentBEIS/inspect_ai}} and DeepEval\footnote{\url{https://docs.confident-ai.com/}} provide partial support for system-level evaluation, they typically concentrate on specific interaction patterns, such as Retrieval-Augmented Generation (RAG) or tool invocations, rather than the full runtime behavior of agents. Effective system-level evaluation must encompass both agent pipelines (including prompts, intermediate results, and final outputs) and artifacts (including goals, memory, reasoning, plans, workflows, tools, knowledge bases, other agents, LLMs, and guardrails)~\cite{dong2024taxonomyagentopsenablingobservability, shamsujjoha2024taxonomymultilayeredruntimeguardrails}. Without such a system-wide perspective, significant failure modes, such as flawed reasoning, incorrect tool usage, or unintended side effects, may remain undetected, undermining the reliability and safety of LLM agents in deployment.

    \item \textbf{Embedding evaluation as a continuous, adaptive process.}
    Unlike traditional software systems, where evaluation is typically confined to predefined testing phases, LLM agents require continuous, adaptive evaluation throughout their lifecycle. These agents update behaviour over time, for example by modifying memory or policies, based on real-world interactions. However, existing evaluation methodologies, such as generalized static benchmarks and predefined test cases, do not adequately capture this evolving nature of LLM agents~\cite{burden2024evaluating,burnell2023rethink}. By incorporating real-time monitoring, retrospective analysis, and structured feedback loops, evaluation can shift from a one-time diagnostic activity to ongoing oversight embedded in everyday operation.

    \item \textbf{Using evaluation results to drive improvements.}
    Continuous evaluation alone is insufficient unless its findings are systematically translated into governed change. In conventional LLM evaluations, performance issues typically lead to model retraining, fine-tuning, or prompt adjustments. For LLM agents, evaluation results must inform a broader range of adaptive changes, including architectural refinements, pipeline or artifact adjustments (for example, operational memory update~\cite{yang2025learning}), and updates to test suites or safety cases~\cite{aisi2024safetycases}. In the absence of structured mechanisms for integrating feedback, adaptations remain fragmented and ad hoc, which limits long-term agent evolution. By embedding explicit feedback pipelines, LLM agents can iteratively evolve to improve decision-making, align with governance requirements, and proactively mitigate emerging risks.
\end{itemize}

To address these challenges, this paper introduces an \emph{evaluation-driven development and operations} (EDDOps) approach that builds on the iterative principles of TDD and BDD while adapting them to the distinctive demands of LLM agents. In this paper, EDDOps refers to the disciplined use of evaluation evidence, both offline and online, to prioritize and govern targeted changes during agent runtime and subsequent (re)development. TDD emphasizes writing automated tests before code changes to provide rapid feedback and guide design at each development step, and BDD structures behavior specifications as executable scenarios to promote a shared understanding of requirements among stakeholders. Similarly, EDDOps embeds continuous feedback loops throughout the entire lifecycle of LLM agents. However, unlike TDD and BDD, which are primarily applied in pre-deployment phases and assume relatively stable specifications and deterministic test outcomes, EDDOps must address the non-deterministic behavior and post-deployment evolution characteristic of LLM agents. In doing so, it positions evaluation as a dynamic and unifying capability that supports ongoing refinement and alignment with evolving objectives and operational contexts.

Guided by a systematic synthesis of evidence, we draw on a multivocal literature review (MLR) to integrate evaluation practices and architectural considerations across the LLM agent lifecycle, capturing insights from both academic and practitioner sources. The study is structured around the following research questions (RQs):

\begin{enumerate}

\item \textbf{RQ1: What are the key challenges in evaluating LLM agents across their lifecycle?}  
To address this question, we conduct a MLR of the current evaluation landscape for LLM agents. The review examines major evaluation dimensions, including contexts, methodological approaches, and architectural concerns, to identify gaps in existing practices.

\item \textbf{RQ2: How can evaluation be \textit{procedurally} integrated into the lifecycle of LLM agents?}  
Building on the MLR data and RQ1 insights, we propose a structured process model that embeds evaluation systematically throughout the lifecycle. The model enables offline and online evaluation, supporting targeted refinements both at runtime and during (re)development.

\item \textbf{RQ3: How can evaluation be \textit{architecturally} integrated into the lifecycle of LLM agents?}  
Informed by the MLR data and RQ1/RQ2 findings, we design a reference architecture that embeds evaluation within the agent’s design. The architecture provides the technical infrastructure needed to support both online and offline evaluation, enabling evidence-driven improvement of the agent.

\end{enumerate}

The remainder of this paper is organized as follows: Section~\ref{sec:background} provides background and related work. Section~\ref{sec:method} details the research methodology including the MLR and the use of empirical evidence from the MLR to derive the proposed process model and reference architecture. Section~\ref{sec:results} reports the MLR results that motivate and parameterize the process model and reference architecture. Section~\ref{sec:ProcessModel} presents a lifecycle process model for evaluation-driven development and operations, and Section~\ref{sec:RefArch} describes a reference architecture shaped by the evidence-derived, architecturally significant requirements. Section~\ref{sec:evaluation} examines the robustness of the MLR and assesses the applicability and adequacy of the process model and reference architecture. Section~\ref{sec:conclusion} concludes and outlines future work.

\section{Background and Related Work}
\label{sec:background}

Evaluation is central to ensuring that LLM agents perform reliably and safely across deployment contexts. Beyond model accuracy, LLM agents require evaluation of reasoning and planning quality, system-level interactions, and dynamic adaptation over time~\cite{xia2024ai}. Evaluation serves multiple purposes, including checking task correctness~\cite{openaioptimizingllm}, examining quality attributes and risks such as fairness~\cite{wang2024beyond,tang2024prioritizing}, and characterizing capabilities and potential unintended consequences~\cite{phuong2024evaluating}. Furthermore, continuous evaluation supports adaptive improvement and risk control by allowing agents and their surrounding systems to refine behavior in response to observed outcomes~\cite{wu2024streambench}. The subsections below situate existing LLM agent evaluation methods, connect them to work that learns from evaluation, and position EDDOps relative to established software development practices.

\subsection{Existing LLM Agent Evaluation Methods}
Existing evaluation frameworks largely focus on the model level~\cite{xia2024ai}, assessing LLM capabilities in areas such as coding~\cite{chen2021evaluating,zhuo2024bigcodebench,du2023classeval} and domain-specific applications (for example, healthcare~\cite{liu2024benchmarking,cai2024medbench}, legal~\cite{fei2023lawbench,guha2024legalbench}, and finance~\cite{xie2023pixiu,islam2023financebench}). While informative, these benchmarks typically assume fixed inputs, static datasets, and aggregate scores, and they often do not account for the system-level sociotechnical interactions that arise in complex, multi-component agents~\cite{liu2023agentbench,gioacchini2024agentquest}.

Many evaluation methods for LLM agents still rely heavily on fixed or periodically updated benchmarks. Static tests risk data contamination~\cite{sainz2023nlp} and do not adapt to the dynamic, evolving contexts of real-world applications. Live or periodically refreshed benchmarks (e.g.,~\cite{white2024livebench,jain2024livecodebench}) mitigate some of these issues by updating tasks over time, yet they generally stop short of enabling in-situ, real-time assessment within the operational environment. As a result, both fixed and periodically updated benchmarks struggle to capture the full complexity and emergent behavior of LLM agents in practical, real-time settings.

System-level evaluation frameworks have received increasing attention, with many approaches emphasizing end-to-end success metrics such as pass rate as the primary measure of performance. Although these metrics provide a holistic view of overall effectiveness, they often overlook critical intermediate decision-making steps that are essential for diagnosing failures and understanding system behavior~\cite{gioacchini2024agentquest}. To address this limitation, step-based and trajectory-based evaluation methods have been proposed~\cite{gioacchini2024agentquest,xiong2024watch}. These methods offer finer granularity, revealing workflow efficiency, agent reasoning, and the progression of decision-making processes. However, overemphasis on step-level or trajectory-level analysis can lead to excessive complexity and make it difficult to aggregate findings into a coherent system-level view~\cite{bentonsabotage}. Effective frameworks therefore need to balance granularity with holistic assessment so that evaluations remain both diagnostically informative and practically actionable.

Another emerging direction is the use of safety cases, which provide structured, evidence-based justifications for the reliability of LLM agents~\cite{buhl2024safety,aisi2024safetycases}. Safety cases help regulators and developers reason about operational risks, but they must be continuously updated to remain aligned with evolving deployment conditions and changing agent behavior. Taken together, these strands indicate substantial progress on evaluation, but they provide limited guidance on how evaluation outputs should systematically drive change in agents and their surrounding systems. This motivates a closer look at work that explicitly uses experience and evaluated evidence to adapt agents.

\subsection{Learning from Evaluation}
A growing body of work seeks to improve agent behavior by leveraging evaluation results: \emph{experience} (such as logged trajectories, future states, and tool feedback) and \emph{evaluated evidence} (such as diagnostics, structured checks, and reward signals). We group the most relevant strands below to highlight how evaluation can feed into adaptation, and to surface the remaining gaps that EDDOps aims to address.

\textbf{Experience-aware reinforcement and replay.}
One strand reuses prior trajectories and preferentially learns from those that provide strong learning signals, such as medium-difficulty, high-information traces rather than trivial successes or hopeless failures~\cite{zhan2025exgrpo,zhang2025rlep}. For example, multiple reasoning or tool-use paths can be sampled and scored with a reward model or verifier, and the model or agent policy is then shifted toward trajectories judged more successful. Replay buffers and fresh rollouts are mixed under stability controls so that the policy improves without collapsing to a single brittle pattern. This line of work mainly operates at development time, offline, as a model- or policy-level intervention before deployment.

\textbf{Reward-free early experience.}
A related line reduces dependence on explicit human reward or dense annotation by treating the agent’s own observed consequences as supervision~\cite{zhang2025agent}. Here, the future state is what actually happened after the agent acted (for example, whether a tool call advanced the task or whether downstream checks passed). By contrasting action–consequence pairs that progressed the task with those that stalled or failed, the agent acquires task-relevant competence that can later seed reinforcement learning, without requiring a curated gold label for every step. This is again primarily a development-time, offline mechanism focused on model- or policy-level adaptation.

\textbf{Environment and curriculum shaping.}
Another strand tunes the training or evaluation environment rather than the agent directly. Tasks are staged from easier to harder, feedback is enriched with informative diagnostics instead of opaque error codes, and dense per-step progress signals are provided; these scaffolds are later removed to verify that behavior generalizes without assistance~\cite{lu2025don}. This treats the evaluation harness itself as an engineering lever during development time and sits at the system or pipeline level, because it modifies interfaces, curricula, and guardrails around the model rather than only tuning the weights.

\textbf{Runtime self-adaptation and self-reflection.}
A further line adapts at runtime, during operation, without redeploying or changing model weights. Reflection-style loops (for example, Reflexion, Self-Refine, and plan–execute–reflect–me\-morize patterns such as MUSE) have the agent critique its own recent behavior, record procedural guidance (for example, ``when you see X, do Y''), and immediately condition later tool calls, routing, or planning on that guidance~\cite{shinn2023reflexion,madaan2023self,yang2025learning}. More recent work extends this with proactive, uncertainty-triggered steering at runtime~\cite{mu2025self}, where the model monitors its own confidence, pauses at high-risk decision points, and corrects course before an error cascades. This improves operational responsiveness but can also introduce drift or unintended side effects if not supervised. This strand is therefore runtime or online and is mostly model- or policy-level.

\begin{table*}
\centering
\caption{Comparison of TDD, BDD, and Evaluation-Driven Development and Operations (EDDOps) for LLM Agents}
\label{tab:tdd-bdd-llm-comparison}
\begin{tabular}{|m{2.6cm}|m{3.4cm}|m{4cm}|m{4.7cm}|}
\hline
\textbf{Aspect} & \textbf{TDD} & \textbf{BDD} & \textbf{EDDOps} \\
\hline
\textbf{Scope} & Pre-deployment (offline)\textsuperscript{*} & Pre-deployment (offline)\textsuperscript{*} & Entire lifecycle (offline and online) \\
\hline
\textbf{Requirements} & Static, explicit & Stakeholder-defined scenarios & Evolving, context-driven \\
\hline
\textbf{Approach} & Unit tests, binary assertions & Acceptance tests, scenario checks & Continuous and adaptive evaluation; both online and offline \\
\hline
\textbf{Feedback Sources} & Developers & Stakeholders & online/offline AI/human evaluators \\
\hline
\textbf{Adaptability} & Low (manual test updates) & Moderate (scenario revisions) & High (evaluation-driven improvements) \\
\hline
\textbf{Validation Focus} & Code-level correctness & Behavior against scenarios & System-level performance and safety \\
\hline
\textbf{Relative Fit} & Lower & Moderate & Higher \\
\hline
\end{tabular}
\\
\begin{flushleft}
\textsuperscript{*}\,In canonical form. Tests may also run in CI/CD after deployment, but the methods are defined around pre-deployment validation.
\end{flushleft}
\end{table*}

\textbf{System-level failure analysis and targeted debugging.}
A final strand moves beyond simply noting that the model produced a wrong answer and treats the system pipeline as the object of diagnosis~\cite{zhu2025llm,yang2025learning}. These efforts collect failing execution traces, classify failure modes (e.g., flawed plan decomposition, wrong tool call, stale retrieval, memory contamination), and identify the earliest critical decision that triggered the cascade. Fixes are then injected where they are most effective: wrapping or throttling risky tools, adding intermediate guardrail checks before irreversible calls, enforcing retrieval constraints, or patching planner prompts at the failure point. In effect, evaluation output is used to alter the runtime pipeline, beyond the base model, so that similar failure modes are less likely to recur. This spans both online and offline: it mines production traces and feeds back into controlled redevelopment, and it is explicitly system-level rather than purely model-level.

Taken together, these strands show that agents can improve by learning from experience and from evaluated evidence. However, most of this work optimizes the agent or its immediate environment in relative isolation, either by updating parameters during development, shaping the training or evaluation environment, or injecting self-authored guidance, memory, or corrective hints during operation. Our work on EDDOps is complementary. We treat evaluation as a first-class system function that spans design time, build time, and runtime operation, where evaluation is not a static leaderboard but a continuous source of evidence, both offline (for example, controlled tests, replay, and scaffolds) and online (for example, live clarification bursts, blocked actions, quarantined results, and drift flags), about where the agent is uncertain, brittle, or overstepping. This in turn calls for a development and operations approach that can systematically absorb such evidence, which we position relative to established practices next.

\subsection{Positioning EDDOps Relative to TDD/BDD}
\label{sec:tdd-bdd-edd}

TDD~\cite{beck2022test} and BDD~\cite{smart2023bdd,enwiki:1250279482} are established methodologies for assuring correctness and stakeholder alignment in software systems with relatively stable, well-defined specifications. TDD encourages writing automated tests before implementation to enforce a red–green–refactor cadence, and BDD validates system behavior through stakeholder-defined scenarios (for example, Given–When–Then) that promote shared understanding.
Table~\ref{tab:tdd-bdd-llm-comparison} contrasts the canonical forms of TDD and BDD with EDDOps.

LLM agents differ in several important respects. They pursue under-specified goals, produce non-deterministic outputs, and continue to adapt after deployment. In their canonical forms, TDD and BDD under-serve these characteristics due to four limitations: (i) reliance on static requirements and executable oracles, (ii) binary pass/fail assertions that do not capture graded or context-dependent outcomes, (iii) a primary focus on pre-deployment validation that can neglect runtime drift and operational feedback, and (iv) limited support for emergent behaviors and qualitative factors such as reasoning coherence, safety, and fairness. These gaps motivate a lifecycle perspective that treats evaluation as an ongoing capability rather than a one-time gate.

EDDOps embeds continuous, adaptive, and actionable evaluation across design time and run time. Grounded in evidence from the MLR, EDDOps treats evaluation as a first-class concern by (i) observing system-level behavior across pipelines and intermediate artifacts, (ii) using quantitative and qualitative signals from human and AI evaluators, and (iii) maintaining feedback loops and traceability so that findings inform iterative refinement of workflows, policies, prompts, reasoning strategies, and tool integration. This approach complements TDD and BDD with evaluation practices suited to open-ended, evolving LLM agent behavior.

\section{Methodology}
\label{sec:method}
\begin{figure*}
    \centering
    \includegraphics[width=0.75\linewidth]{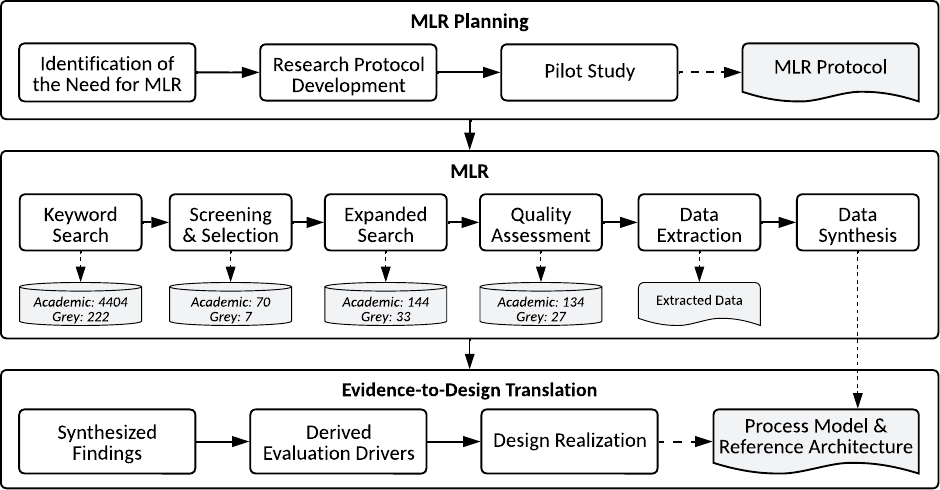}
    \caption{Research Methodology}
    \label{fig:method}
\end{figure*}

This study employs a multivocal literature review (MLR)\footnote{The research protocol and data extraction form can be access here: \url{https://drive.google.com/drive/folders/1VbzxwEHyOJAX1x3k1lIzY2iO64XqmJha?usp=sharing}} (Fig.~\ref{fig:method}) to synthesise insights from both academic and industry sources on LLM agent evaluation, following established guidelines~\cite{Garousi2019,kitchenham2009systematic}.
An MLR is particularly suitable for this topic because evaluation practices for LLM agents evolve rapidly across research and industrial settings, with key methods, tools, and frameworks often documented in technical reports, preprints, or practitioner artifacts.
By integrating these complementary sources, the review captures both theoretical foundations and operational practices that shape how evaluation is conducted in real systems.
Findings from the MLR were distilled into evidence-derived design drivers that informed the development of the process model and reference architecture, providing empirical grounding for both.

\subsection{MLR Planning}
The planning phase identified the need for a multivocal approach to capture both research-driven and practitioner-driven perspectives on LLM agent evaluation.
An initial research protocol was drafted, defining the RQs, search strategy, and inclusion/exclusion criteria.
A pilot review was then used to test the scope and clarity of these elements, ensuring that the protocol captured relevant academic and industrial sources.
Insights from the pilot informed targeted refinements to the search terms and criteria before finalizing the protocol for full data collection.

\subsection{Multivocal Literature Review (MLR)}

\textbf{Databases and Search Strategy}:  
Academic sources were retrieved from Google Scholar, IEEE Xplore, ACM Digital Library, Science Direct, and Springer, while Google Search was used to identify relevant grey literature (e.g., practitioner blogs, evaluation frameworks, open-source tools). The primary search terms were:  
\textit{("large language model" OR "LLM" OR "agent") AND ("evaluate" OR "benchmark" OR "test")},  
with variations (e.g., pluralization, noun-verb shifts, and verb forms) to maximize retrieval. Searches were performed on June 5, 2024, with a focus on literature published since 2022, reflecting post-ChatGPT advancements.

\textbf{Screening and Selection}: After removing duplicates, sources were screened in three stages: title, abstract, and full-text review. Two authors independently conducted each stage to ensure consistency. 
Discrepancies were resolved through discussion. The selection criteria prioritized sources addressing tools, frameworks, or platforms for LLM agent evaluation with clear theoretical or empirical contributions. Excluded sources included those unrelated to evaluation, lacking credibility or substantial contributions, or inadequately documented tools. Eligible sources included academic papers, conference proceedings, technical reports, white papers, and preprints (e.g., arXiv) available in English.

\textbf{Expanded Search}: To broaden coverage and ensure no relevant study was missed, backward and forward snowballing was conducted on the final selection, as per \cite{wohlin2014guidelines}. Additionally, critical new sources published after the initial search were included based on author discussions. Saturation was considered achieved when no significant new themes emerged from additional sources.


\textbf{Quality assessment}.  
To ensure rigor and interpretability, academic sources were screened for credibility (e.g., publication in peer-reviewed venues or authors with relevant research expertise), methodological soundness (e.g., clarity of research design and evidence), and neutrality of interpretation and presentation. Grey literature and tools were assessed for provenance and reliability, considering whether the source originated from recognized organizations or widely used platforms (e.g., GitHub stars), provided sufficient technical documentation, and showed signs of active maintenance or community validation. After this screening, the final dataset comprised 134 academic sources and 27 tools, frameworks, and platforms.

\textbf{Data Extraction and Mapping}:  
Data was extracted using a structured coding framework, mapping key evaluation dimensions relevant to the research questions. This included:
\begin{itemize}
    \item Evaluation Practices: Coverage across pre-deployment, post-deployment, and continuous monitoring.
    \item Evaluation Methods: Types of metrics used (e.g., end-to-end vs. intermediate pipeline/artifact assessments).
    \item Evaluation Outputs and Adaptation Mechanisms: How results inform system refinement, risk control, and iterative improvements.
    \item Architectural Considerations: Evaluation-related components in system design, including feedback loops, observability infrastructure, and safety mechanisms.
\end{itemize}
These insights directly shaped the process model (capturing procedural evaluation integration) and the reference architecture (capturing architectural evaluation integration).


%

\textbf{Thematic synthesis}.  
We conducted a thematic analysis to identify recurring evaluation patterns, challenges, and practices. To complement the qualitative synthesis, we computed frequency counts for the distribution of methods, lifecycle coverage, and adaptation strategies. The resulting themes were directly abstracted into a small set of cross-cutting \emph{design drivers} that summarize recurring evaluation pressures observed across sources. These drivers provided empirical anchors and a shared vocabulary for shaping the process model and reference architecture.

\subsection{Evidence-Informed Derivation of the Process Model and Reference Architecture}
\label{sec:evidence-to-design}

Following the thematic synthesis described above, we related the MLR findings directly to implications for the evaluation lifecycle and for system structure. For each recurring challenge and design driver identified in the synthesis, we examined how existing systems attempted to address or failed to address it, and we abstracted the corresponding procedural responsibilities and structural supports that an evaluation-centric lifecycle and agent stack would need in order to respond to it.

Each implication was captured as a short, actionable \emph{design driver} expressed in plain language (e.g., ``meaningful human oversight''). The authors then used these drivers, together with the underlying coded excerpts, to shape two complementary artifacts:
\begin{itemize}
    \item \textit{Process model shaping}: grouping drivers that impose procedural responsibilities into lifecycle activities and sketching the inputs, outputs, and key steps that explicitly reference evaluation results.
    \item \textit{Reference architecture shaping}: grouping drivers that require persistent structures or interfaces into architectural elements and interactions (for example, evaluators or judges, trace or observability paths).
\end{itemize}

Candidate lifecycle activities and architectural elements were refined through constant comparison with the coded MLR data. Elements were retained or adjusted so that they accounted for multiple observed patterns across sources. In this sense, the process model and the reference architecture emerge as analytical generalizations of the MLR findings, and the design drivers serve as concise labels for clusters of empirical observations.
Consistent with principles for empirically grounded reference architectures by Galster and Angelov~\cite{galster2011empirically}, we treat the MLR as the empirical basis for both the process model and the reference architecture. 

Representative mappings from challenges to design drivers and their realization in the process model and the reference architecture are reported in Section~\ref{sec:results}. The goal of this translation is pragmatic: to keep the process model and the reference architecture visibly tied to empirical evidence, rather than introducing a separate, purely theoretical framework detached from the MLR coding and themes.

\section{RQ1: MLR Results and Takeaways}
\label{sec:results}

To address RQ1, the MLR, comprising 134 academic and 27 grey sources, reveals several critical challenges in current evaluation practices. Percentages are reported with counts; values may not sum to 100 due to rounding. Academic and grey samples are analyzed separately; contrasts are descriptive given the smaller grey sample.

\subsection{Challenges in LLM Agent Evaluation}

\subsubsection{Fragmented Lifecycle Coverage}

We examine evaluation efforts across three primary lifecycle stages: \emph{pre-deployment} (offline (re)development), \emph{post-deployment} (online production or shadow deployment), and \emph{continuous} (spanning both offline and online). On this basis, academic sources skew heavily toward pre-deployment (125/134, 93.28\%), with far fewer addressing post-deployment (3/134, 2.24\%) or continuous evaluation (6/134, 4.48\%). Offline evaluations help validate performance under controlled conditions, but they can miss emergent behaviors, operational degradation, and real-world interactions that arise after deployment.

\begin{figure}
    \centering
    \includegraphics[width=0.6\linewidth]{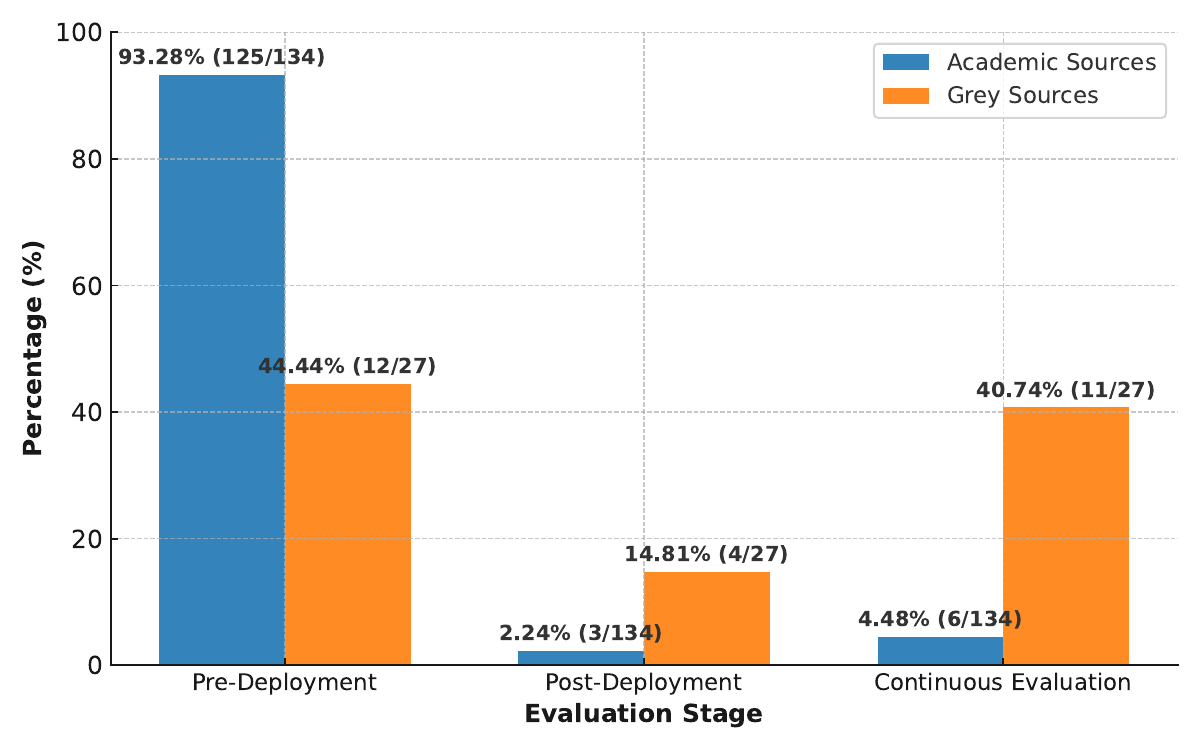}
    \caption{Distribution of Evaluation Efforts Across Lifecycle Stages}
    \label{fig:dis-lifecycle}
\end{figure}

By contrast, grey literature presents a relatively more balanced distribution: 12/27 (44.44\%) pre-deployment, 4/27 (14.81\%) post-deployment, and 11/27 (40.74\%) continuous. This suggests that industry-driven practices recognize the need for ongoing monitoring, though post-deployment evaluation remains underrepresented even in grey sources, indicating that real-world observability mechanisms are still insufficiently integrated.

\textbf{Interpretive takeaways.} The lack of lifecycle-spanning evaluation mirrors technical debt in software engineering~\cite{cunningham1992wycash}: limitations in early-stage evaluation incur operational costs later. In LLM agents, common consequences include:

\begin{itemize}
    \item \textit{Operational Blind Spots.} A pre-deployment emphasis leaves production-only issues (e.g., emergent failures, drift, latency/cost/availability SLO breaches) to go undetected until after deployment, increasing user-visible regressions and shifting mitigation to incident response rather than prevention.
    \item \textit{Broken Feedback.} Siloed stages hinder linking production failures to design changes, slowing targeted remediation.
    \item \textit{Eroding Assurance Over Time.} Sparse post-deployment evidence weakens auditability and the maintenance of safety claims as operating conditions evolve.
\end{itemize}

Addressing this fragmentation requires evaluation frameworks that span the entire lifecycle, aligning with AgentOps principles~\cite{dong2024taxonomy,xia5534588agentops} to support ongoing monitoring, adaptation, and refinement.

\subsubsection{Over-Reliance on Aggregated Metrics}

LLM agent evaluation often reports high-level aggregated outcomes. We classify studies by primary metric type into four mutually exclusive categories: \emph{end-to-end metrics} (single overall scores with no slice- or step-level reporting, e.g., accuracy, task success rate, pass/fail), \emph{intermediate metrics} (measures on specific pipeline stages or artifacts, e.g., planning, tool selection, RAG quality), \emph{mixed metrics} (both end-to-end and intermediate reported as primary results), and \emph{none/not reported} (no explicit metrics, e.g., observability-only reports).

\begin{figure}
    \centering
    \includegraphics[width=0.6\linewidth]{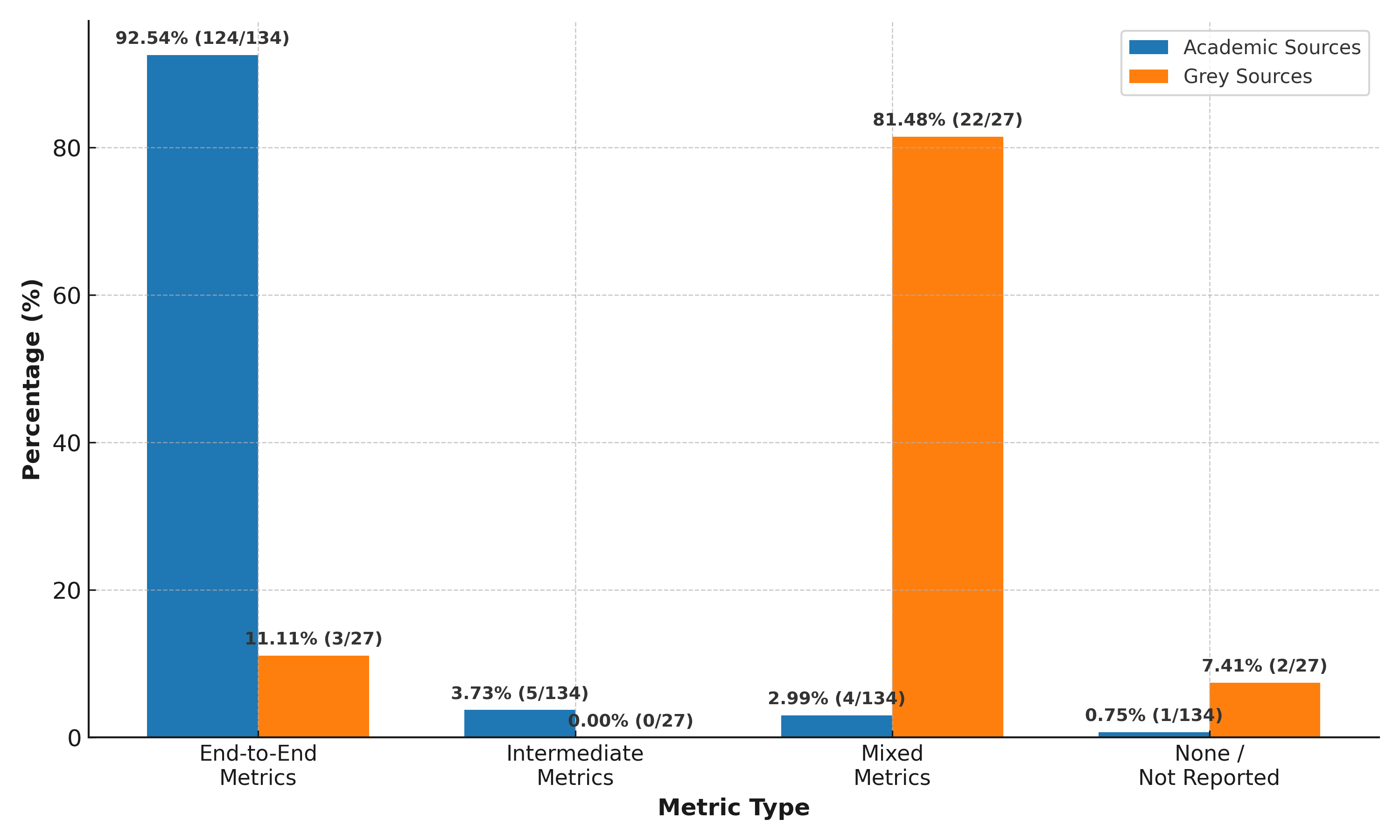}
    \caption{Distribution of Evaluation Metrics}
    \label{fig:dis-metrics}
\end{figure}

As shown in Fig.~\ref{fig:dis-metrics}, academic sources overwhelmingly use end-to-end metrics (124/134, 92.54\%), with very few using intermediate-only metrics (5/134, 3.73\%) or mixed metrics (4/134, 2.99\%); one paper (1/134, 0.75\%) does not report explicit metrics. This distribution shows that purely aggregated outcomes are the dominant primary focus in the academic sample, and that intermediate or mixed reporting is rare.

In the grey literature subset, mixed metrics are most common (22/27, 81.48\%). A small share use end-to-end metrics only (3/27, 11.11\%), none use intermediate-only metrics (0/27, 0\%), and two sources (2/27, 7.41\%) do not report metrics and instead function as observability infrastructure. Many of the tool-oriented grey sources explicitly support configurable metric sets, so end-to-end and intermediate measures can be reported together; none of the tools expose intermediate-only metrics without at least one overall score, which is consistent with their role in operational dashboards.

\textbf{Interpretive takeaways.}
The distributions in Fig.~\ref{fig:dis-metrics} indicate that metric choice has practical consequences for diagnosis and operation. With end-to-end metrics dominating the academic sample and mixed metrics dominating the grey sample, we draw three data-informed interpretations:

\begin{itemize}
    \item \textit{Aggregated-only views limit diagnosis.} When evaluation centers on end-to-end outcomes, the process by which agents reason, plan, and use tools remains largely hidden, which makes root causes harder to localize and increases the likelihood that issues surface late in deployment rather than during routine evaluation.

    \item \textit{Intermediate assessments have trade-offs.} As also reported in \cite{bentonsabotage}, step-level checks face recurring issues: setting meaningful thresholds for narrow capabilities, rolling many fine-grained signals into a coherent system-level view, and higher cost or manual effort for detailed review. The low share of intermediate-only metrics in both samples is consistent with these costs and with the absence of tools that expose intermediate metrics without at least one overall score.

    \item \textit{Mixed metrics balance detail and usability.} The strong preference for mixed metrics in grey sources suggests a pragmatic pattern: pairing targeted intermediate checks with end-to-end outcomes keeps results interpretable and comparable while retaining enough detail for diagnosis. Configurable metric sets allow teams to tie thresholds to overall objectives, summarize fine-grained signals into a small number of indicators, and contain effort by sampling and reviewing only uncertain cases \cite{bentonsabotage}.
\end{itemize}

Taken together, these takeaways motivate treating metric choice as a design decision: end-to-end outcomes provide comparability, but pairing them with selected intermediate checks yields diagnostics that are more actionable for LLM agents.

\subsubsection{Gaps Between Model-Level and System-Level Evaluations}

\begin{figure}
    \centering
    \includegraphics[width=0.6\linewidth]{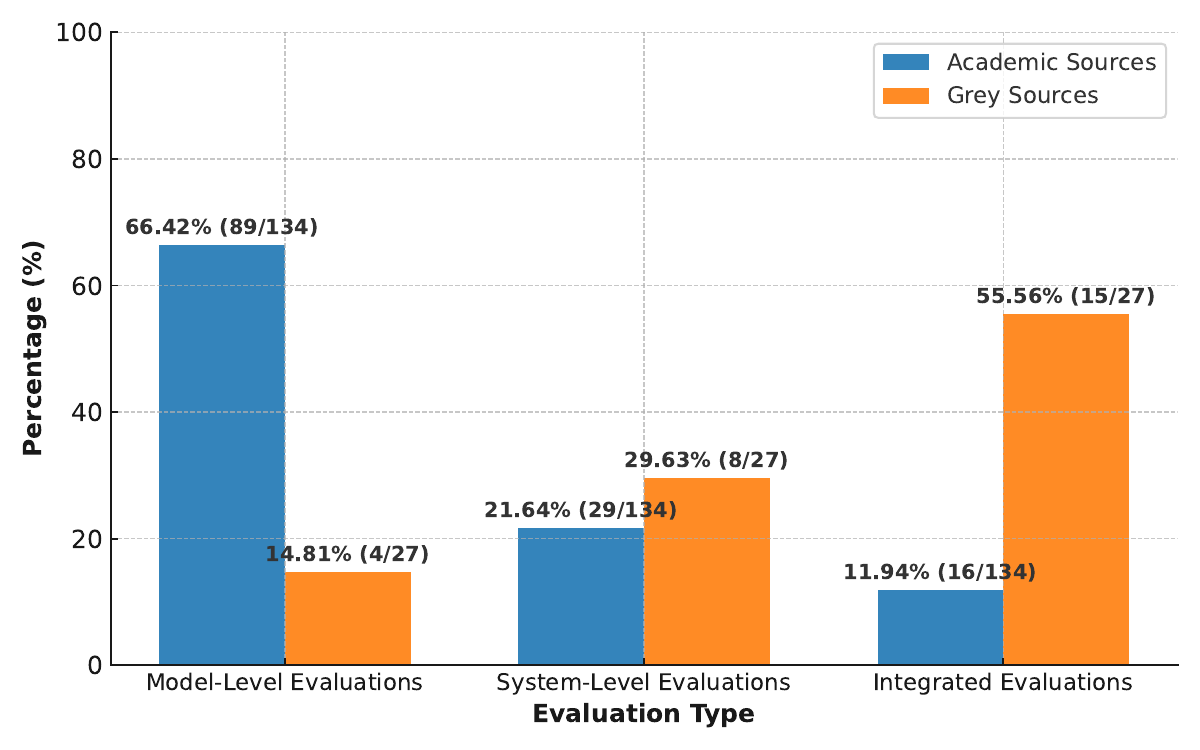}
    \caption{Distribution of Model-Level vs. System-Level Evaluations}
    \label{fig:model-vs-system}
\end{figure}

Current evaluation practice disproportionately focuses on underlying models at the expense of system-level assessments. We classify studies by primary focus into three mutually exclusive categories: \emph{model level} (prompt and response quality or model benchmarks), \emph{system level} (interactions with tools, workflows, or environment), and \emph{integrated} (both levels jointly). As shown in Fig.~\ref{fig:model-vs-system}, academic sources primarily evaluate at the model level (89/134, 66.42\%), with fewer at the system level (29/134, 21.64\%) and a smaller share using integrated evaluations (16/134, 11.94\%). This distribution shows that model level is the most common primary focus in the academic sample and integrated evaluation is the least common.

In contrast, grey literature adopts a more system-aware perspective. In the grey literature subset, integrated evaluations are more common (15/27, 55.56\%), system-level evaluations are 8/27 (29.63\%), and model-only evaluations are 4/27 (14.81\%). This distribution shows that integrated is the most common primary focus in the grey sample and model level is the least common.

\textbf{Interpretive takeaways.}
The distributions indicate that where evaluation is focused shapes which failures are visible and which improvements carry over to operation. We draw three interpretations that align with these patterns:

\begin{itemize}
    \item \textit{Model-only focus can mask system behaviour.} Testing prompts and responses in isolation does not capture orchestration effects, tool behaviour, external dependencies, or error propagation across steps, so models that score well in isolation may still underperform once embedded in workflows.
    
    \item \textit{System-level evaluations surface orchestration faults.} Examining tools, workflows, and environment interactions reveals bottlenecks such as tool call errors, coordination mistakes, and policy enforcement gaps, and it also exposes operational constraints such as latency and cost that model-level tests do not reflect.
    
    \item \textit{Integrated evaluations align fixes with real-world outcomes.} Linking model measures to system outcomes ties reasoning quality to task success and operational objectives, improving traceability from a change in prompts or models to its effect on workflow metrics and user impact.
\end{itemize}

These findings support using the system level as the primary lens for evaluating LLM agents, with model-level probes used to explain and refine observed system behaviour; integrated setups are most useful when they preserve that system-level anchor.

\subsubsection{Lack of Adaptive Evaluations}
\begin{figure}
    \centering
    \includegraphics[width=0.6\linewidth]{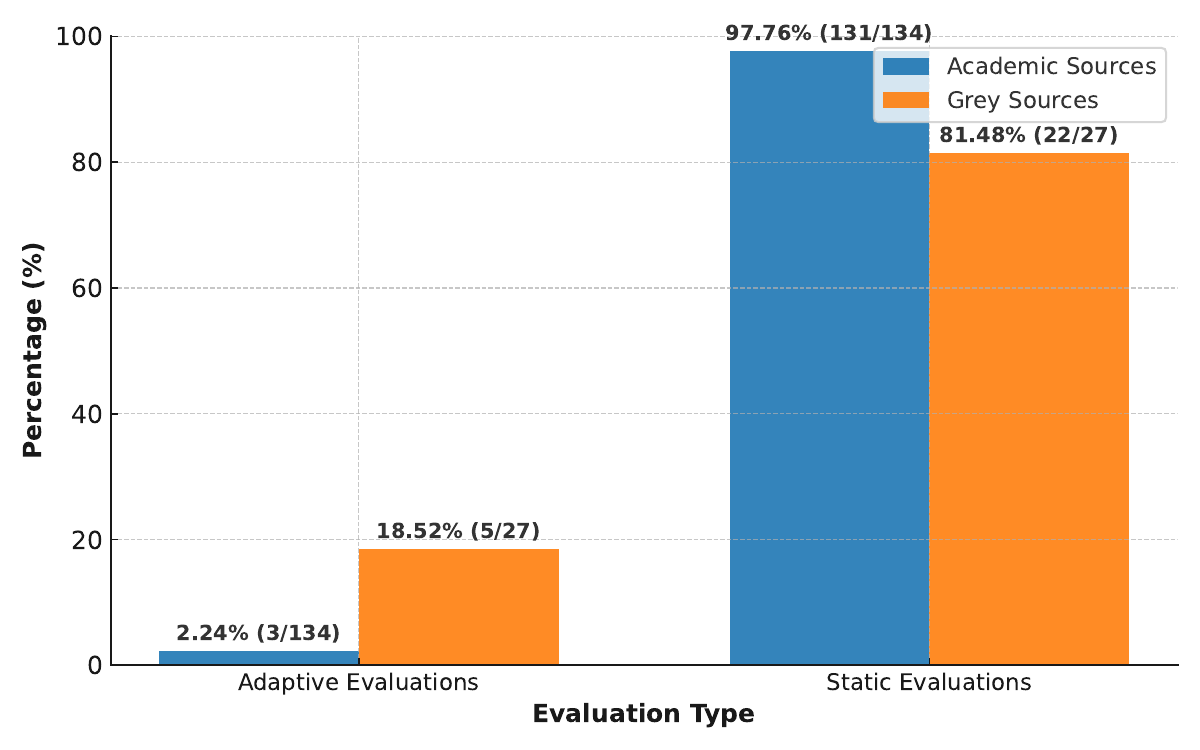}
    \caption{Comparison of Adaptive vs. Static Evaluations in Academic vs. Grey Literature}
    \label{fig:adaptive-evals}
\end{figure}

A common pattern in current LLM agent evaluation is reliance on predefined, static benchmarks and test cases. We classify studies by evaluation mode into two mutually exclusive categories: \emph{static} (fixed suites or predefined benchmarks) and \emph{adaptive} (tests that adjust based on signals such as drift, incidents, uncertainty, or observed failures). As shown in Fig.~\ref{fig:adaptive-evals}, academic sources are predominantly static (131/134, 97.76\%), with very few adaptive (3/134, 2.24\%). 

In the grey literature subset, static evaluations remain the majority (22/27, 81.48\%), while adaptive evaluations are more common than in the academic sample (5/27, 18.52\%). This distribution shows that static is the most common mode in both samples, adaptive is the least common in both, and the share of adaptive is higher in grey than in academic sources.

\textbf{Interpretive takeaways.}
The distributions indicate that choice of adaptive and static evaluation modes affects how well measurements track deployment realities:

\begin{itemize}
    \item \textit{Static Only Checks Miss Real World Variation.} Fixed suites cover a limited slice of possible tasks and inputs, so they are less likely to surface issues that emerge with changing data, tools, or user behavior; coverage constraints increase as task diversity grows~\cite{burden2024evaluating}.
    \item \textit{Adaptive Checks Appear Where Conditions Change.} The higher share of adaptive evaluations in grey sources is consistent with operational settings where inputs, tools, and policies evolve, making some signal driven adjustment useful for keeping tests relevant.
    \item \textit{Balance Stable Baselines With Targeted Adaptive Probes.} Because static evaluations remain the majority while a minority of grey sources employ adaptive methods, a practical pattern is to retain a repeatable static baseline for comparability and add a small number of targeted adaptive checks to increase sensitivity to emerging failures without excessive overhead.
\end{itemize}

In short, a balanced regimen of static baselines and targeted adaptive probes preserves comparability, improves the timeliness of failure detection, and keeps effort bounded as conditions change.

\subsubsection{Failure to Leverage Evaluation for Improvement}

\begin{figure}
    \centering
    \includegraphics[width=0.6\linewidth]{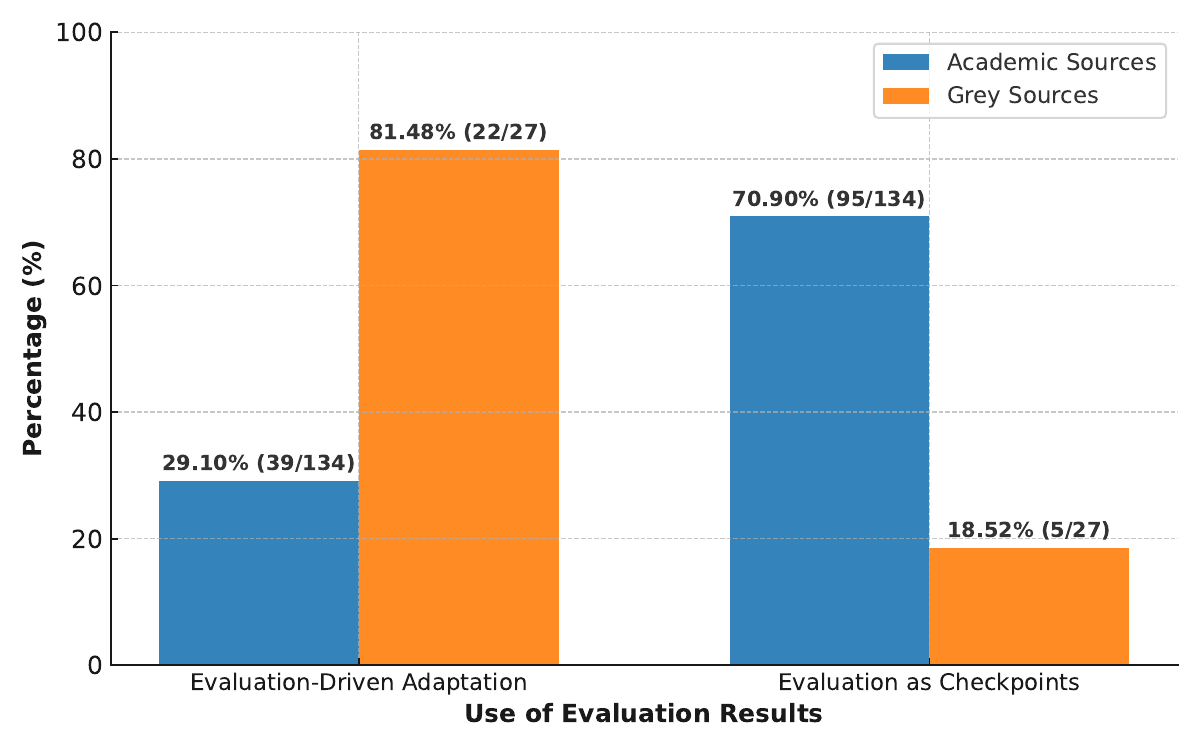}
    \caption{Comparison of Evaluation-Driven Adaptation vs. Evaluation as Checkpoints}
    \label{fig:evaluation-adaptation}
\end{figure}

Although extensive evaluation data is available, its impact on agent refinement is uneven. We classify sources by primary mode: \emph{leverages evaluation} when a study explicitly reports changes made in response to evaluation findings, and \emph{checkpoint only} when evaluation is reported at predefined milestones with no linked changes. As shown in Fig.~\ref{fig:evaluation-adaptation}, in the academic sample 39/134 (29.10\%) leverage evaluation and 95/134 (70.90\%) are checkpoint only.

In the grey literature subset, 22/27 (81.48\%) leverage evaluation and 5/27 (18.52\%) are checkpoint only. This distribution shows that adaptation is the majority pattern in the grey sample, while checkpoint-only reporting is the majority pattern in the academic sample.

\textbf{Interpretive takeaways.}
The distributions indicate that whether evaluation is linked to concrete changes affects robustness and learning over time.

\begin{itemize}
    \item \textit{Open Loops Sustain Repeated Failures.} When evaluation is checkpoint only, findings do not translate into updates to prompts, retrieval, tools, or workflows, so the same issues are more likely to recur in later runs or deployments.
    
    \item \textit{Closed Loops Improve Responsiveness.} When studies report changes tied to findings, evaluation functions as a feedback loop that shortens the path from detection to remedy and helps improvements generalize beyond the specific test.
    
    \item \textit{Tracking Patterns Turns Evaluation Into a Learning Signal.} Recording and revisiting recurring failure types makes it easier to prioritize fixes and to verify that changes address the underlying causes rather than isolated symptoms.
\end{itemize}

In short, linking evaluation findings to explicit changes helps reduce repeated failures and keeps agents responsive as conditions evolve.

\subsubsection{Limitations of AI-Only Evaluations}
We classify evaluator types into three categories by primary mode: \emph{AI evaluators} (automated judging with no documented human review such as LLM-as-a-judge~\cite{zheng2023judging}), \emph{human evaluators} (primary evaluations by humans), and \emph{hybrid evaluators} (combinations of AI and human review). Each source is assigned to the category that best reflects its primary evaluation mode.

\begin{figure}
    \centering
    \includegraphics[width=0.6\linewidth]{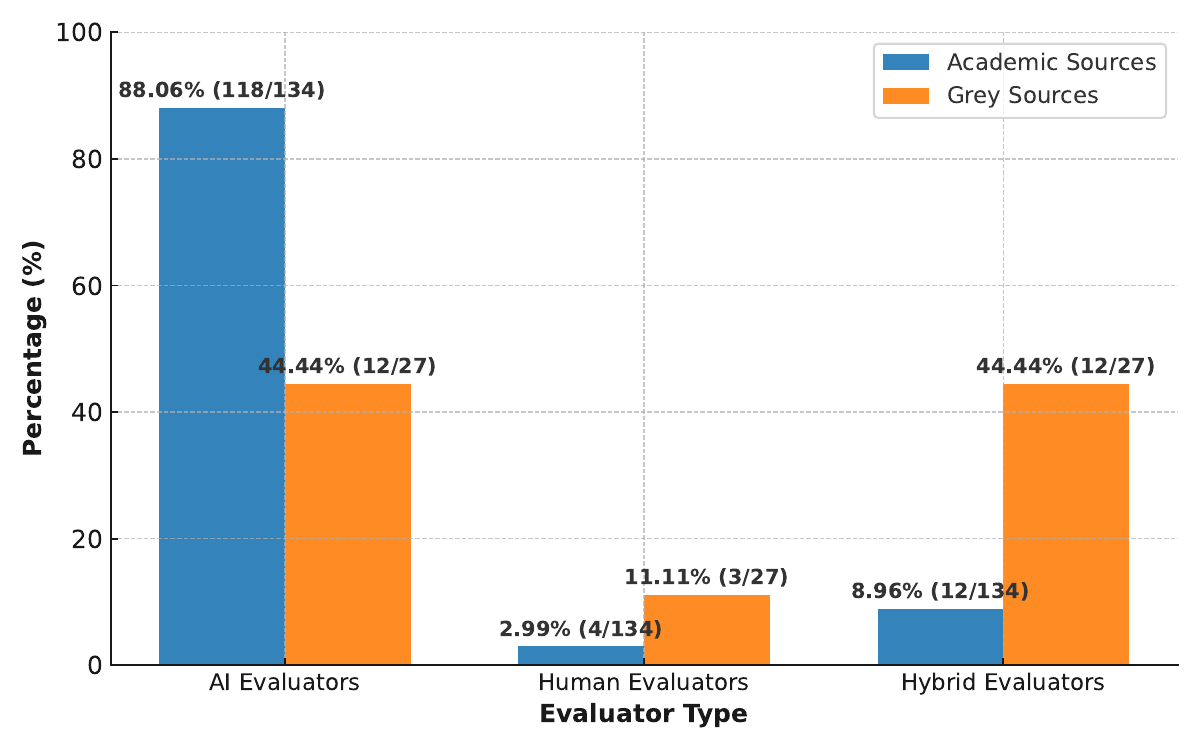}
    \caption{Comparison of AI, Human, and Hybrid Evaluators in Academic vs. Grey Literature}
    \label{fig:ai-human-hybrid-evaluators}
\end{figure}

As shown in Fig.~\ref{fig:ai-human-hybrid-evaluators}, academic sources predominantly use AI-only evaluators (118/134, 88.06\%), with a smaller share using hybrid human–AI evaluators (12/134, 8.96\%) and very few using human-led evaluation (4/134, 2.99\%). This distribution indicates that automated judging is the dominant primary mode in the academic sample.

In the grey literature subset, AI-only and hybrid human–AI evaluators are equally common (12/27, 44.44\% each), and human-led evaluation appears in a minority of sources (3/27, 11.11\%). This distribution suggests a more balanced use of automation and human oversight in the grey sample.

\textbf{Interpretive takeaways.} The observed distributions reveal differing evaluation practices between academic and grey sources. While academic work relies overwhelmingly on AI-only judging, practitioner materials show a stronger preference for hybrid oversight. These patterns suggest several implications for the reliability and governance of LLM agent evaluation:

\begin{itemize}
    \item \textit{AI-only judging can miss context and be miscalibrated.} Prior work shows that automated judges, while scalable, can be less sensitive to reasoning coherence, policy context, and user impact in ambiguous cases, and can inherit bias, display confidence without accuracy, and disagree with human judgements~\cite{ye2024justice,elangovanbeyond}. In light of the strong skew toward AI-only evaluators in the academic sample, these limitations are particularly relevant for LLM agent evaluation.
    
    \item \textit{Hybrid oversight improves reliability where it matters.} The more balanced use of AI-only and hybrid evaluators in the grey sample is consistent with operational needs for meaningful human oversight on uncertain or higher-impact cases. Hybrid setups reported in practice~\cite{zhu2025designing} use humans to review or arbitrate contested or sensitive decisions, which can improve interpretability, catch nuanced errors, and increase confidence in outcomes.
    
    \item \textit{Selective escalation and periodic calibration contain cost.} Several sources describe patterns in which automated evaluation is used by default, with escalation to human review triggered by signals such as uncertainty, disagreement between models, or potential harm~\cite{zhu2025designing}. Periodically calibrating AI evaluators against human ratings helps monitor bias, agreement, and overconfidence over time~\cite{tian2025overconfidence}. This combination preserves much of the efficiency of automation while constraining its risks.
\end{itemize}

In short, heavy reliance on AI-only evaluators risks blind spots, especially for complex or sensitive tasks. Pairing automation with targeted human oversight and routine calibration offers a more reliable basis for evaluating LLM agents while keeping evaluation effort manageable.

\subsection{Cross-Cutting Evaluation Drivers (D1--D6)}
\label{sec:mlr-to-reqs}

Synthesizing across the coded MLR results, we observed a set of recurring patterns that cut across lifecycle stage, metric type, level of focus, adaptation mode, and evaluator type. Rather than introduce a new theoretical layer, we regrouped these patterns into six \emph{evaluation drivers} (D1--D6) that concisely restate what the data repeatedly show: where evaluation tends to be concentrated, where it is missing, and how academic and grey sources differ. 

\begin{itemize}
    \item \textbf{D1 Lifecycle Coverage.} Evaluation should span pre-deployment, post-deployment, and continuous operation, including both offline and online settings.
    \item \textbf{D2 Metric Mix Beyond Aggregates.} Evaluation should combine end-to-end outcomes with intermediate, step-level, and slice-aware checks (for example, by task type or user cohort).
    \item \textbf{D3 System-Level Anchor.} Evaluation should be anchored in system behavior and orchestration, using model-level probes to help explain system-level outcomes rather than evaluating models in isolation.
    \item \textbf{D4 Adaptive Evaluation.} Evaluation should include stable baselines and small, risk- or signal-driven probes that can be adjusted as contexts, tasks, and risks evolve.
    \item \textbf{D5 Closed Feedback Loops.} Offline and online evaluation findings should be linked to concrete actions (such as changes to workflows, policies, or models) and these links should be recorded for traceability.
    \item \textbf{D6 Meaningful Human Oversight.} Evaluation should use hybrid judging patterns so that automated evaluators handle routine cases, while humans retain evaluative agency over ambiguous, high-impact, or contested cases~\cite{zhu2025designing}.
\end{itemize}

Table~\ref{tab:gap_mapping} summarizes how these recurring MLR challenges are reflected in the six drivers and illustrates how each driver is instantiated in the process model and the reference architecture. Combined with the evidence-to-design procedure in Section~\ref{sec:evidence-to-design} which directly leverages the MLR data, this mapping makes the chain from evidence to design transparent: D1--D6 are short labels for empirically grounded regularities, and they are realized procedurally in the process model and structurally in the reference architecture. In this view, evaluation is not a terminal checkpoint but an organizing principle that shapes both development and operation.

\begin{table*}
\centering
\caption{From MLR Challenges to Cross-Cutting Evaluation Drivers (D1--D6) and Their Realization in the Process Model and Reference Architecture}
\label{tab:gap_mapping}
\begin{tabular}{|
>{\centering\arraybackslash}m{3.0cm}|
>{\centering\arraybackslash}m{3.4cm}|
>{\centering\arraybackslash}m{4.6cm}|
>{\centering\arraybackslash}m{4.6cm}|}
\hline
\textbf{MLR Challenge} & \textbf{Evaluation Driver} & \textbf{Process Model (Fig.~\ref{fig:evalflow})} & \textbf{Reference Architecture (Fig.~\ref{fig:RA})} \\
\hline

Fragmented lifecycle coverage &
\textbf{D1 Lifecycle Coverage} &
Step~1 defines lifecycle map; Step~3 conducts offline and online evaluations; Step~4 re-evaluates after changes &
Evaluation Backbone integrates offline/online loops; Operation layer (AgentOps \& Observability) maintains continuous monitoring \\ 
\hline

Over-reliance on aggregated metrics &
\textbf{D2 Metric Mix Beyond Aggregates} &
Step~1 defines metric-mix policy; Step~2 specifies step-level oracles and slice taxonomy; Step~3 reports per-slice results &
Evaluation Repository stores step- and end-to-end results; Agent layer emits fine-grained traces and maintains traceability \\ 
\hline

Gaps between model and system evaluations &
\textbf{D3 System-Level Anchor} &
Step~1 analyzes agent architecture; Step~2 incorporates system considerations; Step~3 exercises full exe. trajectories &
Agent layer exposes evaluation surfaces; Observability standardises schema and IDs \\ 
\hline

Lack of adaptive evaluations &
\textbf{D4 Adaptive Evaluation} &
Step~1 includes risk-based probe triggers; Step~2 generates new tests; Step~3 runs updated tests &
Operation layer triggers probes from Observability; AgentOps tracks online evals \\ 
\hline

Failure to leverage evaluation for improvement &
\textbf{D5 Closed Feedback Loops} &
Step~4 improves the agent online/offline; updates plan in Step~1 and tests in Step~2 &
Evaluation Backbone connects Evaluation Results with Design/Dev Artifacts \\ 
\hline

Over-reliance on AI-only evaluators &
\textbf{D6 Meaningful Human Oversight} &
Step~1 sets escalation policy; Step~3 applies hybrid judging; Step~4 includes human audit of major changes &
Evaluators support AI/human/ hybrid modes; Test and Safety Case Generators embed human checkpoints \\ 
\hline
\end{tabular}
\end{table*}

\section{Process Model for LLM Agent Evaluation}
\label{sec:ProcessModel}
\begin{figure*}
    \centering
    \includegraphics[width=0.9\linewidth]{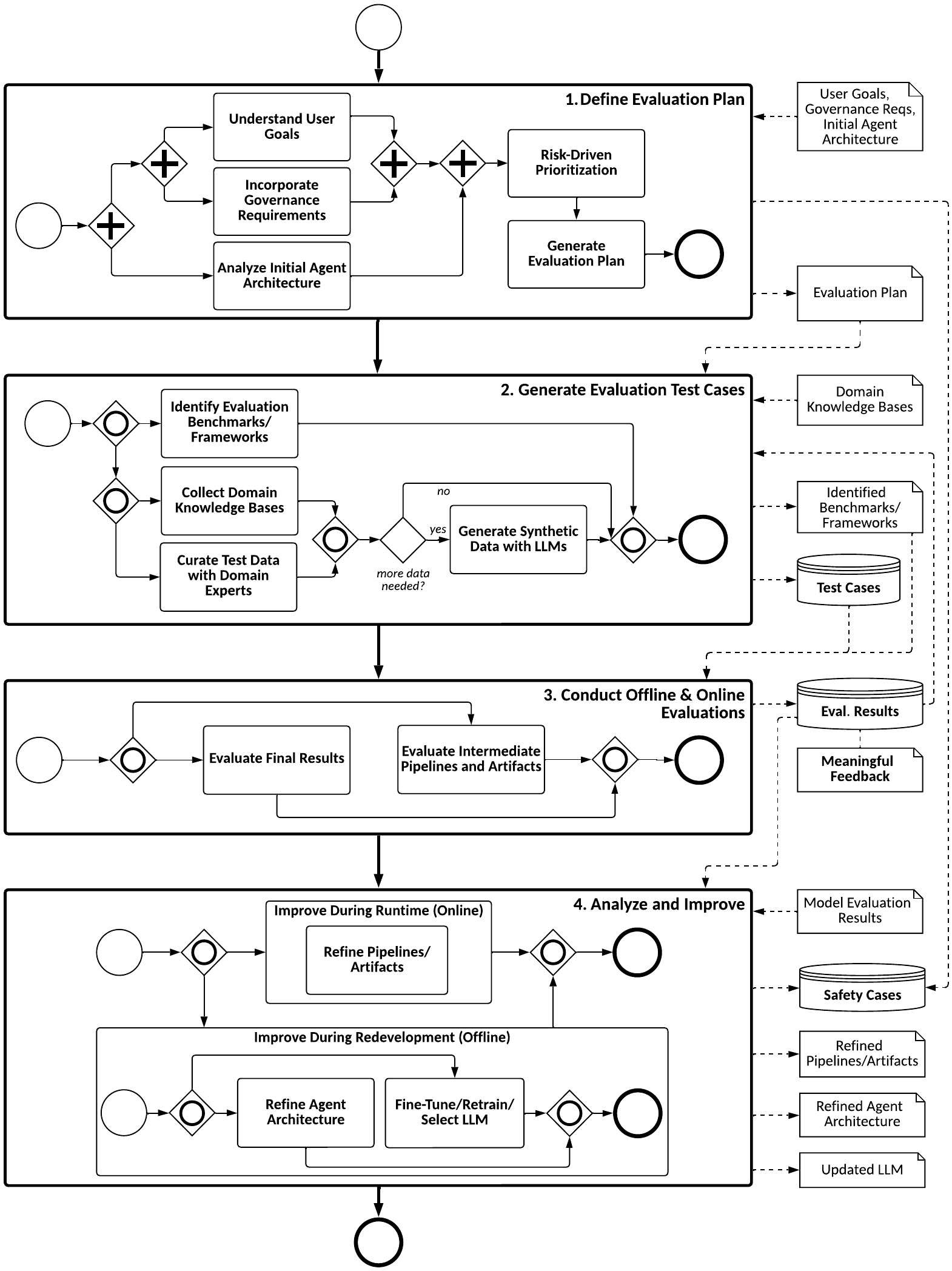}
    \caption{Process Model for LLM Agent Evaluation}
    \label{fig:evalflow}
\end{figure*}

Building on the MLR data and the cross-cutting evaluation drivers (D1--D6), we introduce a structured Process Model for LLM Agent Evaluation (Fig.~\ref{fig:evalflow}). The model reflects common patterns and gaps observed in the MLR and formalizes them into a lifecycle-spanning, risk-aware, and adaptive evaluation process that supports continuous monitoring and refinement across offline and online contexts. It integrates offline and online evaluation, structured feedback loops, and simple human-in-the-loop policies, making evaluation a driver for both development and operations rather than a terminal checkpoint.

\subsection{Step 1: Define Evaluation Plan (D1, D2, D3; enables D4, D6)}

A well-defined plan supports structured, goal-driven, and risk-aware evaluation. Grounded in the MLR findings and the cross-cutting evaluation drivers, Step~1 directly addresses lifecycle coverage, metric mix, and system focus (D1--D3) and sets policy hooks for adaptivity and human oversight (D4, D6). Unlike static, pre-defined plans, this plan is dynamic and lifecycle-spanning: it is updated as user needs, governance, and architecture evolve, with feedback loops and continuous monitoring built in.

\textbf{Key Inputs}: The evaluation plan is informed by three primary inputs:  

\begin{itemize}
    \item \textbf{User Goals}: High level objectives and success criteria that define intended behaviors and contexts (e.g., “An LLM-powered financial advisor should provide tax suggestions tailored to individual income brackets and locations”).
    
    \item \textbf{Governance Requirements}: Legal, ethical, and safety constraints (e.g., ensuring compliance with the EU AI Act) that shape scope and thresholds, and set privacy-safe logging and retention policies for traces~\cite{guldimann2024compl}.

    \item \textbf{Initial Agent Architecture}: The system’s \textbf{structural and behavioral blueprint}, detailing key components, dependencies, and initial risk trade-offs. This input provides insight into technical complexities and potential areas of vulnerability that require focused evaluation.
\end{itemize}  

\textbf{Process Steps}: The evaluation planning process involves \textbf{five key activities} to ensure systematic, risk-aware, and adaptable assessments:

\begin{enumerate}
    \item \textbf{Understand User Goals}: Translate high level goals into structured, testable scenarios with expected interactions, conditions, success criteria, and failure tolerance (e.g., contract drafting with citation and factuality checks)~\cite{beck2022test,adzic2011specification,bartz2020benchmarking}.

    \item \textbf{Incorporate governance requirements.} Carry regulatory and policy constraints into acceptance criteria, evaluator policies, and logging or retention choices (e.g., PII minimization in traces, right to be forgotten)~\cite{guldimann2024compl}.

    \item \textbf{Analyze Initial Agent Architecture}: Identify workflows, tool paths, memories, guardrails, and external dependencies that warrant system level assessment, plus likely failure modes that shape evaluation focus (e.g., RAG latency and tool error handling).

    \item \textbf{Risk-Based Prioritization}: Classify and prioritize scenarios by impact severity (e.g., safety or financial consequences), domain sensitivity (e.g., healthcare versus general use), and stakeholder risk tolerance; set baseline SLO/SLA/SLS targets (latency, cost, quality) as gates; use forward analysis to enumerate plausible risk paths and backward analysis to derive early thresholds from potential harms (e.g., stricter thresholds for medical advice)~\cite{sarmah2024choose,eugeneyan2023llmEvals}.
   
    \item \textbf{Generate evaluation plan.} Consolidate the above into a versioned plan that covers the elements listed in \textbf{Key Outputs} and is aligned with R1–R3 with policy hooks that enable R4 and R6. The plan supports continuous evaluation and iterative improvement via integrated feedback mechanisms so it remains aligned with evolving governance and operating conditions.

\end{enumerate}

\textbf{Key Outputs.} The main output is a structured and adaptive evaluation plan, including:
\begin{itemize}
    \item \textbf{Evaluation Scope and Purpose.} Defines objectives (e.g., accuracy, compliance) and evaluation targets (e.g., reasoning pipelines, retrieval-augmented generation \cite{ru2024ragchecker}, and guardrails \cite{xia2024ai}); sets the lifecycle coverage map across offline, online, and continuous operation, and records prioritized risks and governance constraints that shape scope.

    \item \textbf{Evaluation Strategy and Methodology.} 
    Explain how offline baselines and online checks work together for ongoing validation and monitoring. Specify: (i) the metric mix (end-to-end plus selected step-level checks) with minimum samples per slice; (ii) the run schedule for offline batches and online monitoring; (iii) the evaluator policy, including when AI-only judging escalates to human review; (iv) triggers for adaptive probes (e.g., drift, uncertainty spikes, incidents); and (v) privacy-safe tracing, logging, and retention.

    \item \textbf{Evaluation Criteria and Metrics.} Combines quantitative and qualitative measures (e.g., success rates, response times, robustness scores) with interpretability methods to understand decision processes \cite{hamel2024llmjudge}; defines explicit oracles for step-level checks; and specifies the \emph{slice taxonomy} (e.g., task type, user segment, input source, tool path) and \emph{run metadata schema} (stage labels, slice tags, model or prompt or tool versions, seeds) for traceability, reproducibility, and audit.
\end{itemize}

Additionally, preliminary \textbf{safety cases} may be developed as structured arguments that collect early evidence of safety and compliance; these evolve as evaluation results accumulate \cite{buhl2024safety}.

\textbf{Exit Criteria (Indicators).} The versioned plan includes: a lifecycle coverage map [D1]; a metric mix policy with sampling minima per slice [D2]; a system-level evaluation scope anchored in the current architecture with identified workflows, tools, and step-level signals [D3]; a published slice taxonomy and run metadata schema [D2, D3]; documented adaptive probe triggers [D4]; and a hybrid evaluator policy with escalation criteria and privacy constraints for traces [D6]. Direct: D1--D3; enables: D4, D6.

\subsection{Step 2: Develop Evaluation Test Cases (D2, D3, D4)}

Building on the plan from Step 1, this step translates evaluation objectives into concrete, testable cases. We treat static benchmarks as necessary baselines for initial capability checks, comparability, and regression tracking; at the same time, they are not sufficient for system-level assurance or evolving contexts~\cite{vidgen2024introducing,reuel2024betterbench}. Accordingly, we start from standardized benchmarks and suitable evaluation frameworks, then extend coverage with domain-informed, expert-curated, and synthetic cases that adapt over time. This balances baseline comparability with adaptive depth, addressing the MLR gaps on metric mix (D2), system-level focus (D3), and adaptivity (D4).

\textbf{Key Inputs:} Test case development is informed by three primary sources:

\begin{itemize}
    \item \textbf{Evaluation Plan:} Objectives, scenarios, criteria, slice taxonomy, and run metadata schema from Step 1 ensure alignment with goals, governance, and risk priorities.
    
    \item \textbf{Historical Evaluation Results:} Operational logs and prior findings (e.g., trajectory traces, failure clusters) guide refinement of cases and probes~\cite{langsmith2024evaluation}.
    
    \item \textbf{Domain Knowledge Base:} Domain guidelines, regulations, and expert insights support realistic, high impact cases that reflect operational risks.
\end{itemize}

\textbf{Process Steps:} Test case development follows four interconnected activities designed to ensure both breadth and depth in evaluation:

\begin{enumerate}
    \item \textbf{Identify Evaluation Benchmarks and Frameworks:} 
    \begin{itemize}
        \item \textit{Benchmarks (what to run).} Select standardized benchmarks (e.g., AgentBench~\cite{liuagentbench}) to establish \textit{pinned regression baselines} for core capabilities and cross-version comparability.
        \item \textit{Frameworks/tooling (how to run).} Choose execution frameworks or platforms that match your needs if in-house tools are not available. Ideally they should support: automated runs; step level and trajectory logging; online monitoring; evaluator plugins for AI and human review; privacy-safe logging and export for audit.
    \end{itemize}
     \textit{Example:} for a tax copilot, include RAG and browsing baselines for core tasks and select a framework that captures intermediate artifacts and online metrics.
    
    \item \textbf{Collect Domain Knowledge Bases:} Incorporate authoritative domain-specific sources to create contextually relevant cases~\cite{vidgen2024introducing,reuel2024betterbench,eugeneyan2023llmEvals}. For example, for a tax-copilot deployed in the US, this includes integrating IRS guidelines, tax codes, and relevant case law.
    
    \item \textbf{Curate Test Data with Domain Experts:} Collaborate with subject matter experts to identify edge cases, nuanced scenarios, and potential failure modes not captured by standardized benchmarks \cite{vidgen2024introducing,reuel2024betterbench}. Using the same tax-copilot as an example, tax professionals can help design challenging scenarios, such as resolving conflicting interpretations of tax laws (e.g., home office deductions) or identifying errors in taxpayer-provided data that could trigger audits.
    
    \item \textbf{Generate Synthetic Test Data with LLMs:} Expand coverage where real data are sparse, especially rare or high risk conditions~\cite{long2024llms,huang2024llmmetrics}. This complements, rather than replaces, expert-curated and real-world data.

\end{enumerate}

For each case (benchmark, expert curated, or synthetic), define a step level oracle (pass, fail, or graded) for the relevant pipeline artifacts (e.g., prompts, plans, retrieved knowledge, tool outputs), assign a run mode (offline baseline, online periodic, or signal triggered probe), record provenance (origin and version), and optionally tag slices using the Step~1 taxonomy.

\textbf{Key Outputs:} The outcomes of this step support comprehensive, adaptive evaluation strategies:
\begin{itemize}
    \item \textbf{Versioned Test Catalog:} Identifiers, origins (benchmark, expert, synthetic), slice tags, oracle definitions, run modes and cadence, and version history for traceability and audit.
    \item \textbf{Comprehensive Test Suite:} Risk prioritized coverage that combines pinned regression baselines with domain specific and synthetic cases, exercising both end-to-end outcomes and intermediate artifacts.
    \item \textbf{Selected Benchmarks and Frameworks.} A tailored set of benchmarks and tooling that supports repeatable execution, step level inspection, online monitoring, evaluator plugins (AI and human), slice tagging, and privacy-safe export.
\end{itemize}

\textbf{Exit criteria (indicators).} Pinned regression baselines are defined and documented; the chosen framework supports automated execution, step level logging, slice tagging, online monitoring, evaluator plugins, and privacy-safe export; the suite includes at least one tool or workflow path per critical capability (D3); step level oracles and logging are defined for selected cases (D2); targeted adaptive probes are specified for top risk slices with triggers and cadence (D4); and each case has provenance, slice tags, and an assigned run mode with contamination checks where feasible and a documented leakage policy (train/test/source overlap screening and disclosure) (D2, D4).

\subsection{Step 3: Conduct Offline and Online Evaluations (D1, D2, D3, D4, D6)}

Building on the plan and test catalog from Steps~1–2, this step operationalizes lifecycle coverage (D1), a metric mix beyond aggregates (D2), a system level anchor (D3), adaptive evaluation (D4), and meaningful human oversight (D6). It combines controlled offline evaluation with real-world online evaluation. Offline runs verify that the agent meets baseline standards under controlled conditions. Online runs provide continuous performance and safety monitoring in production-like settings under evolving demands. Together, they form a continuous evaluation and improvement loop that supports adaptation to changing user behaviors, regulatory shifts, and domain constraints.
For clarity, a \emph{slice} denotes a coherent subgroup (e.g., task type, user cohort, operating conditions, or difficulty/uncertainty band) used for targeted analysis and reporting.

\textbf{Key Inputs:} Main inputs to evaluation processes include:

\begin{itemize}
\item \textbf{Identified Benchmarks:} From Step~2, used as pinned baselines for initial capability checks and regression detection, and to inform targeted test case generation.
\item \textbf{Test Cases:} From Step~2, providing routine and edge scenarios that assess both end-to-end outcomes and intermediate artifacts.
\item \textbf{Evaluation Frameworks:} Execution environment selected in Step~2 that can run cases at scale, capture \emph{execution trajectories} (ordered states/actions/tool I/O with timestamps) with slice tags, and support AI and human evaluators across offline and online contexts.\end{itemize}

\textbf{Process Steps.} Evaluation activities center around:
\begin{enumerate}
    \item \textbf{Evaluate final results.} Measure end-to-end outcomes, noting that measures differ by context. \emph{Offline}, teams rely on ground-truth metrics (e.g., task success, exactness, numerical correctness) to establish comparable baselines under controlled conditions; these do not by themselves localize decision faults~\cite{gioacchini2024agentquest,ma2024agentboard}. \emph{Online}, full ground truth is uncommon, so operational proxies are common (e.g., task completion/abandonment, human escalation rate, user satisfaction, latency, unit cost). Online runs use conservative release controls: \emph{shadow evaluation} (0\% user-visible), \emph{canary rollout} to a small traffic share, and staged expansion only when probe and SLO checks pass. Feature flags and a kill-switch enable immediate rollback if anomalies or policy violations are detected. Sampling rates, cost/latency budgets, and per-slice minima are enforced by policy from Step~1.

    \item \textbf{Evaluate intermediate pipelines and artifacts.} Inspect prompts and intermediate results, and analyze execution artifacts such as plans~\cite{valmeekam2024planbench}, retrieved knowledge~\cite{ru2024ragchecker}, and tool outputs~\cite{qu2024tool} to check logical consistency, coherence, and alignment, helping prevent error propagation into final outputs~\cite{xiong2024watch}. Concretely, apply \emph{derived checks} to \emph{future states} (the next observed state after an action), e.g., postconditions, invariants, metamorphic relations, and schema validators. \emph{Offline}, these checks run in replay to localize faults inside trajectories; \emph{online}, lightweight validators and monitors raise slice-level probes when thresholds or disagreement signals are crossed. 
\end{enumerate}

Teams may use AI judges for scale~\cite{zheng2023judging,vertex_ai_eval_metrics}, complemented by human review where ambiguity, stakes, or low-confidence signals warrant it~\cite{aisi_joint_testing_exercise_2025}. When AI judges are used, basic assessor-health indicators (uncertainty/dispersion summaries; simple consistency checks) are recorded per slice before decisions are acted upon.

\textbf{Outputs:} This step generates a set of evaluation results, offering complementary insights:

\begin{itemize}
\item \textbf{Offline Evaluations:} Baseline metrics, error analyses, and pass/fail summaries from controlled scenarios for regression tracking; \emph{post-mortem} scoring of captured trajectories with defect localization (enabled by trajectory capture).
\item \textbf{Online Evaluations:} Real-world measures including user impact, adaptive responses, and behavioral patterns under operational variability; per-slice anomaly flags; and \emph{inline} validator results over future states (e.g., “no error banner,” schema invariants) attached to trajectories for later replay and analysis. These same validators can gate the distillation of procedural/strategic/tool memories at test time~\cite{yang2025learning}.
\item \textbf{Actionable Feedback:} Synthesized findings (affected slices, exemplar trajectories, proposed probes) that populate tickets or change requests for Step~4.
\end{itemize}

\textbf{Exit Criteria (Indicators).} Results are labeled as offline or online and include slice identifiers and full provenance (D1, D2, D3). Minimum per-slice samples for the reporting window are met (D2). Anomaly signals trigger probes within target latency and outcomes are logged (D4). The evaluator policy is applied with recorded escalation decisions (D6).

\subsection{Step 4: Analyze and Improve (D5; enables D1--D4, D6)}
This step translates evaluation evidence into change (D5), allowing the agent to learn from its own past experience~\cite{zhan2025exgrpo,zhang2025agent}. Using the results from Step~3, it applies targeted improvements through (i) bounded, policy-approved runtime adjustments and (ii) governed offline redevelopment, ensuring human oversight and traceability are preserved (D6). These actions build directly on the lifecycle coverage, metric-mix diversity, system-level focus, and adaptive feedback mechanisms established in earlier steps (D1–D4). By \emph{targeted}, actions are confined to the specific subset of cases that underperformed, hereafter a \textit{slice} (a coherent subgroup defined by task type, user cohort, operating conditions, or difficulty/uncertainty band)~\cite{zhan2025exgrpo}. The \textit{smallest effective change} is prioritized—the least intrusive intervention expected to resolve the observed failure (for example, a prompt or routing adjustment before architectural or model modification). All changes are versioned, auditable, and traceably linked to their originating findings.

\textbf{Key Inputs:}
\begin{itemize}
    \item \textbf{System-level evaluation results (from Step~3):} Offline (controlled) and online (operational) findings, which may be disaggregated by slice and labeled with slice identifiers (e.g., task type, user group, difficulty, uncertainty band)~\cite{acikgoz2025self}. Results may include quantitative metrics and qualitative notes that surface baseline gaps, emergent behaviors, and user-visible issues, together with evaluator-health metadata (e.g., agreement rates, uncertainty/confidence) to help avoid acting on noisy aggregates.
    \item \textbf{Model evaluation results:} Although the focus is system-level, vendor reports and public benchmarks can reveal model tendencies; these can inform system-level mitigations (e.g., guardrails, prompts/policies, routing) and, where justified, model changes.
\end{itemize}

\textbf{Process Steps:}
\begin{enumerate}
\item \textbf{Improve during runtime (online).} Bounded adjustments may be applied to the affected slice(s) under pre-set limits, with logging for auditability and rollback. Illustrative actions include:
\begin{itemize}
    \item selecting or minimally adjusting a prompt/profile for a specific intent that appears to fail under time pressure;
    \item serving conditional in-context examples when an intent detector matches the failing pattern;
    \item altering routing or policy for the slice (e.g., directing scheduling requests to a stronger toolchain; tightening retry/backoff);
    \item modifying guardrail thresholds for a sensitive domain cohort; or writing evaluation-gated operational memories (procedural/strategic/tool) for the affected intent~\cite{yang2025learning}.
    \item updating tool configuration (e.g., shorter timeouts or safer fallbacks when an upstream API drifts).
\end{itemize}

\item \textbf{Improve during redevelopment (offline).} Where issues exceed runtime bounds or recur, governed design changes can be introduced and verified offline before broader use. Typical actions include:
\begin{itemize}
    \item \textbf{Refining architecture/specification.} Strengthening planning scaffolds, retrieval/data paths, and tool contracts; tightening acceptance criteria where evaluations reveal ambiguity~\cite{abuelsaad2024agent}. For example, adding precondition checks before invoking destructive tools or reconfiguring retrieval filters for date-sensitive queries.
    \item \textbf{Updating the model (when justified).} If deficits are attributable to model capability, considering targeted fine-tuning/retraining with selected previous evaluation results~\cite{zhang2025agent} or selecting a better-suited model~\cite{christiano2017deep,ge2023openagi}; pairing any model change with system-level mitigations (e.g., post-generation verification) to maintain holistic reliability.
\end{itemize}
After redevelopment, re-run the relevant \emph{offline} evaluations on the \emph{same slices} to confirm the intended deltas; pooled aggregates can mislead. For \emph{online} adaptation (updates during live interaction), rely on continuous inline validation with guardrails (e.g., allow memory writes only after a checklist pass) to prevent ``misevolution''~\cite{shao2025your}.
\end{enumerate}

\textbf{Outputs:}
\begin{itemize}
\item \textbf{Safety cases:} Updated safety cases that link changes to risk claims and controls with slice-level evidence~\cite{buhl2024safety}.
\item \textbf{Refined agent architecture:} Documented modifications to components and specifications aligned with observed gaps and operational goals.
\item \textbf{Updated LLM (if applicable):} Fine-tuned, retrained, or newly selected models, with stated limitations and associated mitigations.
\item \textbf{Refined pipelines/artifacts:} Adjusted prompts, policies, tools, workflows, memory artifacts, and, where appropriate, conditional in-context example sets addressing the failing intents.
\end{itemize}

\textbf{Exit Criteria (Indicators).} Each prioritised finding is mapped to a concrete change entry; the associated post-change checks are defined and executed on the affected slices (D5, D4). Evaluator reliability is satisfactory. Where a change affects risk claims or controls, the safety case is updated with the new evidence and rationale (D5, D6).

\section{EDDOps Reference Architecture} 
\label{sec:RefArch}

\begin{figure*}
    \centering
    \includegraphics[width=0.9\linewidth]{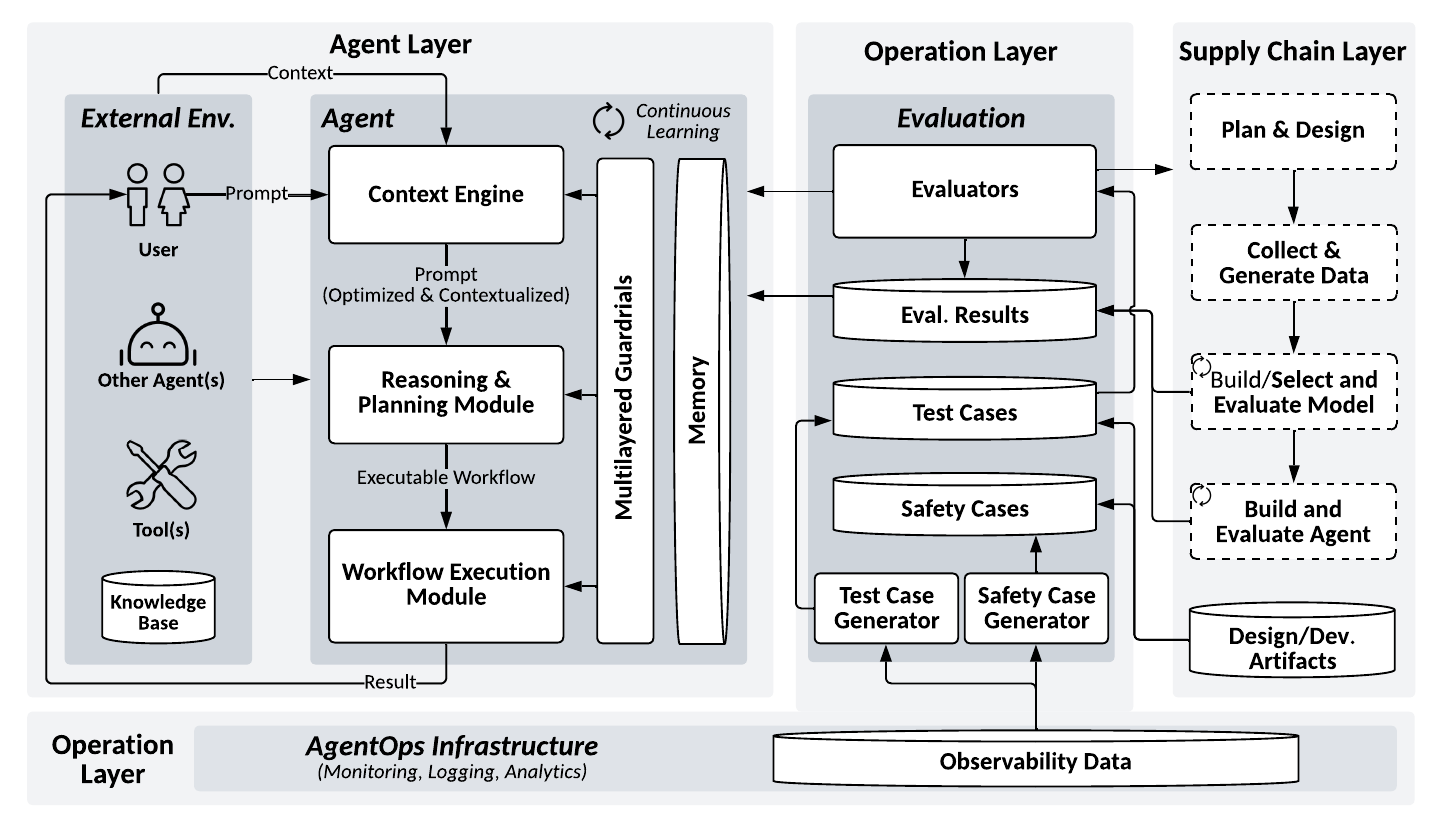}
    \caption{Reference Architecture for Evaluation-Driven Development and Operations (EDDOps). The Evaluation Backbone and Control Loop links evidence to bounded runtime adaptations and governed redevelopment.}
    \label{fig:RA}
\end{figure*}

To address RQ3, we propose a reference architecture (Fig.~\ref{fig:RA}) that analytically generalizes the MLR findings and the cross-cutting evaluation drivers (D1--D6), and complements the process model from Section~\ref{sec:ProcessModel}. Consistent with principles for empirically grounded reference architectures~\cite{galster2011empirically}, the architecture is positioned as an industry-cross-cutting, preliminary facilitation reference architecture for multiple organizations: it abstracts from domain-specific details while remaining tied to the coded evidence and design drivers described in Section~\ref{sec:evidence-to-design}. Candidate components and control flows were iteratively checked against the MLR themes and drivers, and retained only when they accounted for recurrent patterns observed across sources.

The architecture adopts an \emph{evaluation-first} stance: an \textbf{Evaluation Backbone and Control Loop} connects offline and online evaluation to drive changes during development and operations (D1--D5), with hybrid oversight (D6), rather than treating evaluation as a terminal checkpoint. Concretely, the backbone embeds lifecycle coverage (D1), a metric mix beyond aggregates (D2), a system-level anchor (D3), adaptive evaluation (D4), closed feedback loops (D5), and meaningful human oversight (D6) as first-class responsibilities.

Our design is guided by three key principles, grounded in patterns observed in the MLR:

\begin{itemize}
    \item \textbf{Lifecycle integration} (D1). Evaluation spans pre-deployment and post-deployment, enabling continuous visibility and improvement across the agent lifecycle.
    \item \textbf{Evaluation as a driver of change} (D5). Findings instantiate bounded runtime adjustments and governed redevelopment, recorded as evidence-linked changes rather than static reports.
    \item \textbf{System-anchored, adaptive, and overseen} (D2, D3, D4, D6). Mixed metrics with step-level signals, system-level anchoring of evaluations, signal-triggered probes, and hybrid human/AI oversight.
\end{itemize}

Existing reference architectures for foundation model\-based agents~\cite{lu2024towards} and responsible AI systems~\cite{lu2023responsible} acknowledge evaluation only peripherally or as isolated validation stages. Our architecture instead centers \emph{evaluation-as-control}, extending prior work with a single, evaluation-centric responsibility grounded in D1--D6: an \textbf{Evaluation Backbone and Control Loop} that causally links evidence to both development-time and operation-time changes (D1--D5) and supports governed human oversight where needed (D6). This backbone moves evaluation from a passive logging activity to an active driver of refinement and assurance across the \emph{Supply Chain}, \emph{Agent}, and \emph{Operation} layers, operationalizing the evaluation drivers as architectural capabilities rather than optional add-ons.

\subsection{Supply Chain Layer}

This layer establishes evaluation intent and the evidence base for downstream evaluation (D1--D3). It is not a complete development reference architecture; it highlights those supply-chain activities that are shaped by evaluation needs and that feed the Evaluation Backbone:

\begin{itemize}
    \item \textbf{Plan and design}. Define user goals, governance constraints, and a high-level architecture, and derive the evaluation plan from Step~1 of the process model. These artifacts are created explicitly to structure later evaluation and traceability, not only implementation.

    \item \textbf{Collect and generate data}. Prepare datasets for pinned baselines and slice-aware checks in line with the evaluation plan (Step~2 of the process model), recording provenance and basic contamination/leakage screens.

    \item \textbf{Build/Select and evaluate model}. Choose candidate models and run initial checks (e.g., system cards, black-box tests) against the planned metrics, noting limitations that system-level mitigation and evaluation must cover (Step~3 of the process model).
    
    \item \textbf{Build and Evaluate System}. Integrate model, tools, context, memory, and guardrails, and run offline system-level evaluations to generate initial trajectories, failure modes, and safety arguments before any live traffic (Step~3 of the process model).
\end{itemize}

The \emph{Design/Dev.\ Artifacts} repository aggregates these outputs as versioned architecture descriptions, evaluation plans, seed test cases, and draft safety cases (D2,D3,D5). Through Step~4 of the process model, later offline and online evaluation results feed back into this store so that subsequent design and development changes remain grounded in accumulated evidence rather than ad-hoc judgment.

\subsection{Agent Layer}

This layer hosts the adaptive core but, from an EDDOps perspective, its primary role is to expose \emph{system-level evaluation surfaces} and to emit well-structured traces that the Evaluation Backbone can consume (D2,D3,D5). Boundaries and interfaces are drawn so that reasoning, tool use, and guardrails are observable and controllable, not only functionally correct.

\textbf{External Environments}. Users (feedback and impact signals), other agents (coordination)~\cite{chen2023agentverse}, tools/APIs (capabilities), and knowledge bases (domain/context) act as sources of observable consequences that can be recorded and evaluated . These environments are treated as explicit evaluation contexts (e.g., user cohort, tool family, knowledge source).

\textbf{Agent Module}. Core components are instrumented to support evaluation:
\begin{itemize}
    \item \textbf{Context Engine}. Aggregates inputs from the environment and memory and applies slice and run metadata tags so that subsequent decisions can be analyzed by specific operating conditions.

    \item \textbf{Reasoning \& Planning}. Generates plans and workflows and emits plan-level traces (identifiers, steps, tool intents) suitable for step-level or trajectory-level evaluation and localization of flawed decomposition or planning.
    
    \item \textbf{Workflow Execution}. Executes tool calls and actions and outputs execution traces (inputs, outputs, status, latency) that allow derived checks and probes to target specific tools, paths, and slices.
    
    \item \textbf{Multi-layered Guardrails}. Apply safety and policy checks with logging and escalation hooks~\cite{shamsujjoha2024taxonomymultilayeredruntimeguardrails}, enabling hybrid evaluation and human oversight for high-risk actions.

    \item \textbf{Memory}. Stores task and feedback context, including evaluation-informed operational memories, so that adaptations can be explained, audited, and, if required, rolled back.
\end{itemize}

Supply-chain artifacts such as test cases, safety cases, and model or system cards provide expectations against which these traces are checked. Evaluation results from the Operation Layer feed back into this Agent Layer by adjusting prompts, routing, guardrail settings, and memory policies within bounded limits (D4,D5,D6). In this way, the Agent Layer is both the object of evaluation and the locus where evaluation-driven changes are applied under governance.

\subsection{Operation Layer}

This layer is the center of continuous evaluation and adaptation. It operationalizes the \emph{Evaluation Backbone and Control Loop} (D1--D6) by connecting (i) offline evaluation results from the Supply Chain Layer, (ii) online signals from observability, and (iii) governed change paths back to engineering and runtime configuration. In effect, it makes evaluation both the main source of evidence and the main driver of improvement.

\subsubsection{Evaluation (Backbone and Control Loop)}
The \textbf{Evaluation Backbone} integrates evidence produced across the lifecycle and routes it through two complementary paths:

\textbf{Offline Improvement Path (development-time)} (D1, D2,D3,D5).  
Batch evaluations (e.g., benchmarks, curated suites) produce evidence stored in \emph{Evaluation Results}. These results flow into \emph{Design/Development Artifacts}, supporting updates to system artifacts and pipelines like prompts, tool interfaces, planning logic, guardrails, and test/safety cases. Each change should be versioned and linked to its originating evidence so that re-evaluation closes the loop before redeployment.

\textbf{Online Adaptation Path (operation-time)} (D1,D2,D4, D6).  
Runtime evaluations, triggered by observability signals or scheduled probes, deliver live feedback to \emph{AgentOps}. Within predefined governance boundaries, these results can adjust prompts, routing policies, or guardrail thresholds without redeployment; higher-impact changes are deferred to the offline path. Hybrid evaluation policies determine when automation suffices and when escalation to human review is required.

\textbf{Evidence Stores and Traceability} (D2,D3,D5,D6).  
Versioned \emph{Evaluation Results}, \emph{Test Cases}, and \emph{Safety Cases} form the knowledge base that both paths rely on. Their structure (e.g., schema IDs, slice tags, workflow/tool identifiers) enables reproducible re-runs, targeted regression checks, and auditable evidence$\rightarrow$change relationships.

\subsubsection{AgentOps Infrastructure and Observability}

The AgentOps Infrastructure is the operational environment where evaluation becomes actionable (D1--D6). It provides the mechanisms for collecting structured traces, triggering additional checks, and enacting decisions produced by the Evaluation Backbone. Observability components emit standardized logs, including plan traces, execution traces, slice tags, stage labels, and workflow or tool identifiers. These logs support both online monitoring and offline batch evaluation.

Operational controls enable evaluation to steer behavior:

\begin{itemize}
    \item \textbf{Release Controls} (D1,D4,D5).  
    Shadow evaluation on mirrored traffic, canary rollout, and staged expansion allow new configurations to be exercised safely. Feature flags and kill-switches support rollback when evaluation evidence indicates regressions.

    \item \textbf{Proactive Error Analysis} (D2,D4).
    Metrics, anomaly signals, and trace patterns trigger slice-targeted probes during operation. The same observability logs also support offline error analysis, such as clustering failures by slice, reconstructing trajectories, and inspecting tool paths. This keeps evaluation sensitive to both drift observed at runtime and recurring issues discovered through retrospective analysis.

    \item \textbf{Feedback Integration Pipeline} (D5,D6).  
    Probe results, escalation outcomes, and offline error analyses are routed into engineering workflows as tickets or configuration updates. Evidence-to-change relationships are preserved for audit, oversight, and scheduled recalibration of evaluators.

\end{itemize}

Overall, the AgentOps Infrastructure ties together online evaluation, offline log-driven analysis, and evaluation-informed operational controls. It ensures that the system remains continuously observable and continuously improvable.

\section{Empirical Validation}\label{sec:evaluation}

This section provides a \emph{fitness-for-purpose} validation of our contributions. 
Because the process model and reference architecture are empirically derived directly from the MLR, the goal here is not to re-establish their empirical basis but to confirm that they are \emph{useful and adequate} for their intended purpose. 
Following Galster and Angelov~\cite{galster2011empirically}, this constitutes an \textit{analytical and instantiation-based} validation rather than an experimental one: the artifacts are processual and architectural, so quantitative metrics (e.g., task accuracy) are not directly applicable, and validation instead focuses on applicability, adequacy, and coherence. 
We therefore triangulate along three lenses: 
(i) the \textbf{methodological soundness} of the MLR (threats to validity), 
(ii) the \textbf{applicability} of the process model through a caselet and practitioner synthesis, and 
(iii) the \textbf{architectural adequacy} of the reference architecture through comparative and analytical checks.

\subsection{Methodological Soundness of the MLR}

\textbf{Internal validity.} Explicit inclusion/exclusion criteria, double screening with reconciliation, and a documented coding protocol reduce selection and synthesis bias. Practitioner read-throughs were used to sanity-check major themes for plausibility in real settings.

\textbf{External validity.} The corpus spans academic (n = 134) and grey (n = 27) sources. Transferability to specific domains depends on context, so the MLR serves as a \emph{derivation base}; applicability is then confirmed through the caselet and practitioner triangulation reported below. Broader multi-domain replication is future work.

\textbf{Construct validity.} Concepts such as lifecycle coverage, metric mix, and system-level anchoring are operationalized as cross-cutting evaluation drivers (D1–D6). Derivation tables (Table~\ref{tab:gap_mapping}) preserve traceability from findings to these drivers, and all extraction sheets and codebooks are archived for reproducibility and critique.

\subsection{Applicability of the Process Model}

We evaluate whether the process model is understandable and actionable for planning, running, and closing evaluation loops. 
Evidence comes from (i) a pre-deployment caselet using an LLM-based tax assistant and (ii) synthesis of publicly documented practitioner experiences.

\subsubsection{Caselet: LLM-based Tax Assistant}

The model was instantiated in a pre-deployment setting for an LLM-based tax assistant (as described in~\cite{LU2026112656}) that provides citation-backed guidance over Australian Taxation Office materials~\cite{lu5266496agentarceval}. 
The exercise covered planning, offline evaluation, and improvement proposals (no production rollout).

\textbf{Step 1 (Plan).} Goals (reliable, citation-backed advice), governance constraints, and high-level architecture were defined. 
Risk-based prioritization focused on high-impact slices e.g., ambiguous deduction rules). 
Outputs included a metric-mix policy, slice taxonomy, run metadata schema, and escalation policy.

\textbf{Step 2 (Prepare test cases).} A neutral evaluation framework (DeepEval) with step-level logging was selected, and domain cases were curated with subject-matter input (for example, home office deduction, cross-border income). 
Synthetic augmentation targeted edge conditions only and was documented with leakage checks.

\textbf{Step 3 (Evaluate).} Offline batches were run to measure end-to-end and step-level indicators (e.g., retrieval hit rate, plan coherence). 
Human review audited failures, and operational proxies (latency/cost budgets, escalation rates) were specified, though not exercised in production.

\textbf{Step 4 (Improve).} Findings produced evidence-linked tickets for prompt or query adjustments and guardrail refinements. 
Runtime controls and rollback criteria were proposed and versioned as part of a change plan.

In this caselet, the model provided a clear path from plan to tests, evaluation, and change, with traceable artifacts and decisions. 
All evaluation drivers (D1–D6) were instantiated in some form in this scope; online controls (shadow/canary) remain to be exercised under actual deployment conditions.

\subsubsection{Practitioner Triangulation}

Practitioner sources~\cite{hamel2024llmjudge,huang2024llmmetrics,cognition2024review,hamel2025evalQA} confirm that mechanisms aligned with the process model are already used in practice and map closely to its main steps:

\textit{Interpretable plans and clear oracles.}  
Practitioners emphasise simple pass/fail criteria and explicit rationales for outcomes, aligning with the model’s focus on explicit oracles, metric mix, and slice design.

\textit{Evolving benchmarks, not static suites.}  
Teams extend baselines using production logs, error clusters, and user feedback, mirroring the model’s adaptive test growth and signal-triggered probes.

\textit{Paired offline/online evaluation.}  
Offline runs anchor comparability under controlled conditions, while online proxies and light auditing supply real-time signals for operational slices.

\textit{Hybrid oversight with calibration.}  
Ambiguous or high-impact cases are routed to humans, and AI–human agreement is tracked over time, with thresholds adjusted as drift appears.

\textit{Closed-loop improvement and guarded rollout.}  
Findings trigger evidence-linked tickets, feature-flagged runtime changes, and rollback checks before broader exposure.

Taken together, the caselet and practitioner synthesis indicate that the process model is both understandable and directly applicable to current LLM-agent engineering practice, while remaining consistent with the evaluation drivers distilled from the MLR.

\subsection{Architectural Adequacy of the Reference Architecture}

We assess whether the architecture’s allocation of responsibilities realizes the evaluation drivers D1–D6 and improves on related architectures. 
Validation follows two complementary approaches: (i) comparative analysis against prior Responsible-AI and FM-agent architectures, and (ii) analytical evaluation against ISO/IEC 25010 quality attributes.

\subsubsection{Comparative Analysis}

Our architecture extends prior Responsible AI and FM-agent designs~\cite{lu2023responsible,lu2024towards} by making \emph{evaluation a first-class responsibility}. 
Table~\ref{tab:comparative_analysis} summarizes the main differences. 
In earlier architectures, evaluation appears as one component or pattern (e.g., Continuous RAI
validator), whereas here it is treated as a structural control backbone that realizes D1–D6 across lifecycle and operation.

\begin{table*}
\caption{Comparative analysis highlighting evaluation as a structural, first-class concern.}
\centering
\label{tab:comparative_analysis}
\begin{tabular}{|
>{\centering\arraybackslash}m{2cm}|
>{\centering\arraybackslash}m{3.5cm}|
>{\centering\arraybackslash}m{3.8cm}|
>{\centering\arraybackslash}m{5cm}|}
\hline
\textbf{Dimension} & \textbf{Responsible AI Systems Design \cite{lu2023responsible}} & \textbf{FM Agent Design \cite{lu2024towards}} & \textbf{EDDOps Architecture (this work)} \\
\hline
Scope & Responsible AI principles for system design & Foundation-model-based autonomous agents & Evaluation-driven development and operations of LLM agents \\
\hline
Modularity & Compliance and governance modules & Autonomy-oriented, coordination modules & Pluggable evaluators, repositories, and observability controls \\
\hline
Evaluation integration & Continuous RAI
validator pattern for audits & Continuous assessments for risk/governance & Evaluation Backbone linking offline/online loops \\
\hline
Observability & Audit logs and dashboards & Task monitoring and self-reflection & Full observability via AgentOps instrumentation \\
\hline
Human-in-the-loop & Governance and audit checks & Decision refinement & Hybrid oversight for meaningful HITL \\
\hline
\end{tabular}
\end{table*}

\subsubsection{Analytical Validation}

We analytically validated whether the responsibilities in the reference architecture support the ISO/IEC~25010 quality characteristics that matter for evaluation–driven operation. The architecture does not guarantee these qualities by itself, but it allocates the concrete mechanisms needed to realize them in an implementation.

\textbf{Functional suitability.}  
The Evaluation module, including Evaluators, Eval.~Results store, Test Case Generator, Safety Case Generator, and Safety Cases, together with the feedback paths to the Supply Chain and Agent layers, provide a complete set of functions for planning, executing, and using evaluations, rather than ad hoc checks scattered across components.

\textbf{Performance efficiency.}  
AgentOps Infrastructure and the Observability Data store centralize monitoring, logging, and analytics, so evaluations can be scheduled, sampled, and scaled independently of the agent logic. This allows teams to tune evaluation frequency and depth to respect latency and resource constraints while still collecting sufficient evidence.

\textbf{Compatibility.}  
The architecture separates the Agent from External Environments (users, other agents, tools, knowledge bases) through explicit interfaces, and routes all evaluation through a shared Evaluation component. This layered structure supports interoperability across different tools, models, and deployment platforms.

\textbf{Interaction capability.}  
Although user interfaces are out of scope, the architecture exposes evaluation artifacts in a form that is consumable by people: sliceable Eval.~Results, explicit Test Cases, and structured Safety Cases linked to Design/Dev.~Artifacts. This organization supports recognizability and operability for engineers, reviewers, and auditors who need to inspect and act on evaluation outcomes.

\textbf{Reliability.}  
Multilayered Guardrails, Memory, and the closed-loop connections from Eval.~Results back to the Agent and Supply Chain layers support fault detection and controlled recovery. When evaluations reveal problems, bounded online improvement and offline improvement paths provide structured mechanisms to change behavior without destabilizing the system.

\textbf{Security.}  
Guardrails enforce safety and policy checks on agent actions, while Eval.~Results, Safety Cases, and Design/Dev.~Artifacts are treated as versioned evidence stores. This supports integrity and accountability of evaluation data and of the risk arguments built on top of it.

\textbf{Maintainability.}  
The architecture is modular: Evaluation, AgentOps Infrastructure, Agent components, and Supply Chain activities are separated into distinct responsibilities, and key artifacts (tests, results, safety cases, design documents) are explicitly represented. This modularity and versioning support analyzability, regression testing, and safe evolution of evaluation logic and agent behavior.

\textbf{Flexibility.}  
The three-layer structure and pluggable Evaluation block allow the same architecture to be instantiated across domains and traffic levels. Different evaluators, test generators, and safety-case templates can be plugged into the same backbone without changing the overall structure.

\textbf{Safety.}  
Risk-aware planning in the Supply Chain layer, Multilayered Guardrails in the Agent layer, and the Safety Case Generator and Safety Cases in the Operation layer provide explicit mechanisms for risk identification, argumentation, and safe integration of changes informed by evaluation.

Taken together, these checks indicate that the reference architecture allocates responsibilities in a way that supports the key quality characteristics relevant to evaluation-driven development and operation, and that it is adequate as an empirically grounded, facilitation-oriented reference architecture rather than a prescriptive implementation template.

\section{Conclusion}
\label{sec:conclusion}

This paper presented an empirically grounded approach, EDDOps, for embedding evaluation as a continuous, governing function throughout the lifecycle of LLM agents. Guided by three research questions, it advanced both understanding and practice for evaluation-driven development and operations. Through a MLR (RQ1), we identified recurring limitations in existing evaluation practices, including fragmented lifecycle coverage, over-reliance on aggregate metrics, limited adaptivity, and insufficient human oversight, and we distilled these findings into cross-cutting evaluation drivers. These drivers, along with the MLR data, directly constrained and informed two artifacts: a process model (RQ2) that procedurally embeds continuous evaluation by linking planning, evaluation, and improvement through traceable feedback loops, and a reference architecture (RQ3) that structurally realizes evaluation as a backbone connecting offline and online loops, supported by observability, hybrid oversight, and evidence-based change control. Taken together, the process model and reference architecture position evaluation not as a terminal checkpoint but as a persistent system capability that drives adaptive refinement, sustains safety and accountability, and supports the responsible evolution of LLM agents in dynamic environments.



\bibliographystyle{unsrt}  
\bibliography{refs}

\end{document}